\documentclass[pra,preprint,superscriptaddress,floatfix,showpacs, 11pt]{revtex4-1}
\usepackage{hyperref} % For hyperlinks
\usepackage{amsmath,amsfonts,amssymb}
\usepackage{graphicx}
\usepackage{xcolor}
\usepackage{makecell}
\usepackage{natbib}

\begin{document}
\title{Dynamics and transport of Bose-Einstein condensates in bent potentials}
\author{Rhombik Roy}
\email{rroy@campus.haifa.ac.il}
\affiliation{Department of Physics, University of Haifa, Haifa 3498838, Israel}
\affiliation{Haifa Research Center for Theoretical Physics and Astrophysics, University of Haifa,
Haifa 3498838, Israel}
\author{Ofir E. Alon}
\affiliation{Department of Physics, University of Haifa, Haifa 3498838, Israel}
\affiliation{Haifa Research Center for Theoretical Physics and Astrophysics, University of Haifa,
Haifa 3498838, Israel}

%\date{\today}

\begin{abstract} 
The dynamics of bosons in curved geometries have recently attracted significant interest in quantum many-body physics. Leveraging recent experimental advances in tailored trapping landscapes, we investigate the quantum transport of weakly interacting bosons in two-dimensional bent trapping potentials, showing that geometry alone can serve as a precise control knob for tunneling dynamics. Using time-adaptive many-body simulations, complemented by mean-field analysis and exact diagonalization, we analyze both static and dynamical properties of bosons confined in the bent potential. We reveal how bending an initially straight channel induces a transition from density localization to delocalization and drives the buildup of correlations in the ground state. In the dynamics, the bent acts as a tunable barrier that enables controllable tunneling: weak curvature allows coherent tunnelling across the bend, while stronger bent suppresses transport and enhances self-trapping. The tunneling rate can be precisely tuned by geometric parameters, establishing bent traps as versatile platforms for geometry-controlled quantum transport. 
\end{abstract}

\maketitle

\section{Introduction}\label{intro} 
The study of quantum transport in engineered geometries has become a central theme in the quest to control and manipulate quantum matter~\cite{refb1,refb2,refb3,refb4}, making it a unifying theme across various areas of physics~\cite{PhysRevLett.134.207102,Hartnoll2015,Chou_2011,KOHLER2005379}.
Among these, bent or curved potentials offer a particularly rich platform, where the geometry itself acts as a control knob for quantum dynamics~\cite{refbb1,refbb3}. Such systems reveal intriguing effects, ranging from geometrically induced bound states to non-trivial transport and interference phenomena, that have no counterpart in straight, symmetric geometries~\cite{refbe1,refbe2}. Bent geometries have garnered increasing interest across diverse fields, including quantum waveguide design~\cite{refbd1}, cold atom trapping~\cite{refbd2}, mesoscopic electron transport~\cite{refbd3}, and atomtronics~\cite{refbd4}, where curvature and asymmetry can be harnessed to control flow or induce localization.

Bose-Einstein condensates (BECs) confined in curved geometries provide a versatile platform for studying how spatial curvature and topology shape the behavior of quantum matter. The geometry-induced modifications to the energy, boundary conditions, and effective potentials can significantly alter key properties such as coherence, excitation spectra, and transport dynamics. For example, shell-shaped condensates realized in bubble traps allow for controlled tuning between filled and hollow geometries, leading to notable shifts in the critical temperature and the nature of collective excitations~\cite{ref2b1, ref2b1a}. Moreover, curvature can influence the formation and stability of topological excitations such as vortices, whose dynamics become intertwined with the underlying geometry~\cite{ref2b2,ref2b3,ref2b4}. These developments open pathways to investigate curvature-driven quantum phases and nonequilibrium transport in low-dimensional systems.

In this paper, we investigate the transport properties of bosons in a two dimensional bent potential, highlighting how the curvature introduces new possibilities beyond those offered by standard flat or one-dimensional potentials. Using the many-body numerical framework, we examine how geometric features influence key dynamical phenomena, such as tunneling and depletion.
Our motivation stems from recent experimental advances that enable the realization of curved trapping geometries for ultracold atoms. In particular, techniques like painting potentials with moving laser beams allow for the creation of toroidal, ring-shaped, and bent configurations~\cite{bent_exp1,bent_exp2}, enabling controlled manipulation and transport of BECs in complex landscapes~\cite{bent_exp3,bent_exp4}. 
We focus on a class of geometries generated by bending a straight trapping channel, which gives rise to local minima and rich spatial structures~\cite{S0129055X95000062,PhysRevE.87.042912}. These bent potentials go beyond the limitations of one-dimensional systems, offering additional degrees of freedom for tailoring transport dynamics---without relying solely on varying potential barriers, as is typical in traditional trapping potentials. 

Our study investigates both the static and dynamical properties of bosons confined in the bent potential. In the first part, we analyze the ground-state properties and demonstrate that by solely tuning the geometry of the bent potential, effective local minima emerge. The many-body ground-state analysis reveals that increasing the bend width causes the one-body density to gradually separate into two distinct regions and the system transitions from condensation to fragmentation. Key factors such as the sharpness of the bent, the width of the curved region, and the steepness of the transverse confinement also play critical roles in controlling the density localization and many-body fragmentation. Thus, we show how geometrical aspects of trapping with no lower-dimension counterparts govern the ground state properties.

Building on this controlled realization of localization of the density, the second part of the paper explores the dynamical behavior, specifically the transport through the bent potential and the mechanism to control the transport. We show that tunneling dynamics can be finely tuned by solely changing the bent geometry, offering a level of control not available in standard potentials. Our results, thus, establish the bent potential as a versatile platform for studying the transport phenomena using both mean-field and many-body methods. 

The structure of this paper is as follows. In Section \ref{system}, we introduce the physical system, describe the form of the bent potential, outline the many-body framework employed, and define the key observables. Section \ref{results} presents our numerical results on both static properties and transport behavior in the bent geometry. Section \ref{conclusion} summarizes the main findings and provides concluding remarks. 
In Appendix~\ref{apndx}, we further examine the low-lying energy spectrum of non-interacting bosons in the bent potential using exact diagonalization. These results are then analyzed to identify the relevant two-mode or multi-mode contributions, providing deeper insight into bosonic transport in the bent geometry. The supplementary material includes additional analysis of the transport dynamics, along with the convergence of the many-body results.

\section{System}\label{system}
The properties of $N$ interacting bosons---both in their ground state and during time evolution---are described by the many-body  Schr\"odinger equation $\hat{H} \Psi = i\frac{\partial \Psi}{\partial t} $.
The Hamiltonian of the system is given by
\begin{equation} \label{hamiltonian}
\hat{H} = \sum_{i=1}^{N} \hat{h}(\mathbf{r}_i) + \sum_{i<j} \hat{W}(\mathbf{r}_i - \mathbf{r}_j),
\end{equation}
where the one-body operator is defined as $\hat{h}(\mathbf{r}_i) = -\frac{1}{2} \frac{\partial^2}{\partial\mathbf{r}_i^2} + V_{\text{trap}}(\mathbf{r}_i),$
with $\mathbf{r} = (x, y)$ represents the position in two spatial dimensions.  $V_{\text{trap}}(\mathbf{r})$ denotes the external bent trapping potential, which is described in details below. 
The two-body interaction is modeled by a finite-range Gaussian potential:
$\hat{W}(\mathbf{r}_i - \mathbf{r}_j) = \frac{\lambda_0}{2\pi \sigma^2} \exp\left[-\frac{(\mathbf{r}_i - \mathbf{r}_j)^2}{2\sigma^2}\right]$,
where $\lambda_0$ is the interaction strength and $\sigma = 0.25$ is the interaction width~\cite{sudip_asymmetric_dw,anal_2020_dw}. The effective interaction parameter is defined as $\Lambda = \lambda_0 (N - 1) = 0.1$. 
It is important to note that, in the mean-field framework, both ground-state and dynamical properties depend solely on the interaction parameter  $\Lambda$. Consequently, the particle number $N$ can be chosen arbitrarily for a given fixed $\Lambda$. 

We solve the  Schr\"odinger equation in dimensionless units. The dimensionless form is obtained by scaling the Hamiltonian with the factor $\hbar^2 / (mL_0^2)$, where $m$ is the boson mass and $L_0$ is a characteristic length scale~\cite{MCTDHB2,rhombik_acc}.  In this representation, energies are expressed in units of $\hbar^2 / (mL_0^2)$, lengths in units of $L_0$, and time in units of $mL_0^2 / \hbar$.

\subsection{Bent Potential} \label{bent_pot}
\begin{figure}
    \centering
    \includegraphics[width=0.75\textwidth, angle =-0 ]{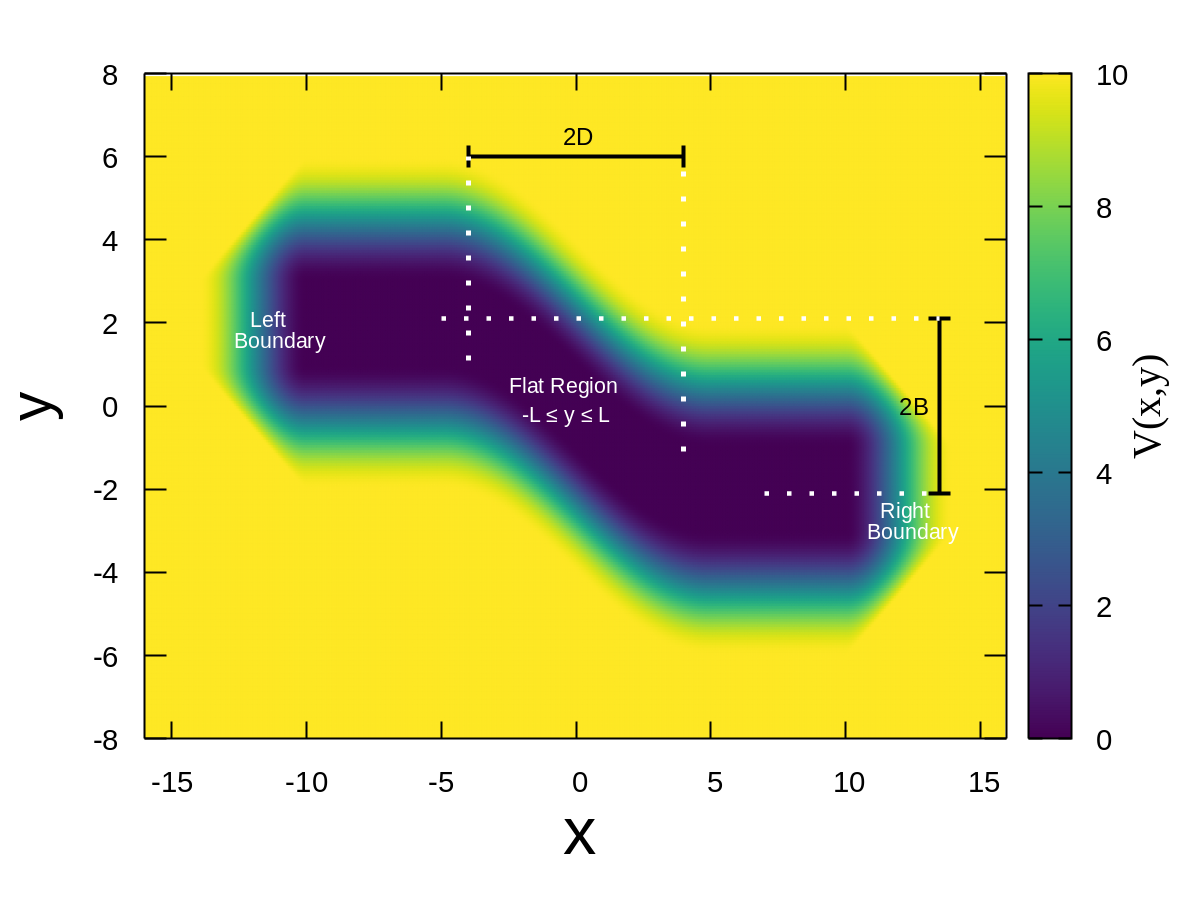}
    \caption{Schematic diagram of the bent potential $V(x,y)$. The potential has the total width $2B$ and extends a distance $2D$ along the x-direction, which sets the sharpness of the bent. The central flat region spans $-L < y < L$.  Boundaries are smoothly closed along the x-direction to prevent bosons from escaping. See text for more details. The quantities shown are dimensionless. }
    \label{fig0}
\end{figure}

We investigate the transport of bosons in the two-dimensional bent potential, $V(x,y)$. This potential is formed by introducing a smooth curvature along the x-direction combined with transverse confinement in the y-direction. The curvature is introduced by modifying the longitudinal coordinate as, 

\begin{equation}\label{potentialx}
    y= 
\begin{cases}
    y-B,& \text{if } x < -D, \\
    y-2B sin \left(\frac{\pi (x-D)}{4D} \right)^2 + B ,& \text{if } -D \leq x \leq +D ,\\
    y+B ,& \text{if }  x> +D. \\
\end{cases}   
\end{equation}
The parameter $B$ controls the vertical shift of the bent and is referred to as the half-width of the bent, while $D$ determines the extent of the bent region in the x-direction and sets its sharpness, see Fig~\ref{fig0}. Confinement along the $y$-direction is provided by a soft-rising potential:

\begin{equation}\label{potentialy}
    V(x, y) = 
\begin{cases}
    V_0, & |y| > L + a, \\
    V_0 \sin^2\left( \dfrac{\pi(|y| - L)}{2a} \right), & L \leq |y| \leq L + a, \\
    0, & |y| < L,
\end{cases}
\end{equation}
where $V_0$ is the maximum potential height, $2L$ defines the width of the central flat region, and $a$ controls the smoothness of the transverse confinement, see Fig~\ref{fig0}. Throughout the calculations and analysis, we fix the parameter $L=0.5$.  Together, Eqs.~\eqref{potentialx} and~\eqref{potentialy} define a curved channel that guides the condensate during the transport. The written bent potential is opened in the x-direction. But during the computational analysis, to stop the bosons from escaping, the two ends are closed smoothly by ramping up the barrier. 
The functional form of the right and left barriers is  $\left[V_0 \sin^2\left( \dfrac{\pi(x \pm 8)}{2a} \right)\right]$. The schematic diagram of the bent potential is shown in Fig.~\ref{fig0}. 
The potential is defined within a box of size $[-16,16)$ in both the $x$- and $y$-directions, discretized using $128 \times 128$ grid points. 

We divide our investigations into three sections. In the first part, the ground state properties of the trapped bosons in the bent potential is studied. In the second part, we analyze the transport properties of the bosons in the bent potential and the final part discusses how the transport in the bent potential can be controlled by varying different geometric parameters.

\subsection{Methodology and observables}
We use the multiconfigurational time-dependent Hartree method for bosons (MCTDHB) to solve the $N$-boson Schr\"odinger equation~\cite{MCTDHB1,MCTDHB2}. This method expands the many-body wave function using time-dependent permanents, created by distributing $N$ bosons across $M$ single-particle orbitals~\cite{mctdhb_exact2} 
\begin{eqnarray}
\vert \Psi(t)\rangle = \sum_{n} C_{n}(t)\vert n;t\rangle; \hspace{5ex}
\vert n;t\rangle = \prod_{i=1}^{M}
 \frac{ \left( b_{i}^{\dagger}(t) \right)^{n_{i}} } {\sqrt{n_{i}!}} \vert vac \rangle .
\label{many_body_wf}
\end{eqnarray}
The system comprises of $M$ orthonormal time-dependent single-particle states, denoted by $\phi_j (x,t) =\langle x\vert \hat{b}_j^{\dagger}(t) \vert vac \rangle$, where $ b_{j}^{\dagger}(t)$ is the creation operator.
MCTDHB yields exact results when the chosen set $\vert n;t \rangle$ spans the complete Hilbert space, which occurs in the limit $M \rightarrow \infty$. For computational efficiency, we select a finite value of $M$ in the numerical simulations, ensuring it is sufficiently large to achieve highly accurate results. 
The number of grid points are also important in this study to describe the bent accurately.  In the supplementary material, we show the convergence of our numerical results in terms of time-adaptive orbitals as well as grid points.
The time-dependent variational principle is employed~\cite{variational5}, leading to two coupled set of equations that are solved simultaneously to determine the time evolution of the wave function~\cite{mctdhb_software1,mctdhb_software2}. For a more detailed explanation and applications, see the review~\cite{mctdhb_review} and the related work~\cite{rhombik_scirep,paolo_cavity2,paolo_cz,rhombik_pre,fischer_Metrology,GWAK2021168592,rhombik_quantumreports,paolo_PhysRevA.98.053620,paolo_cavity,rhombik_epjplus,paolo_mott,fischer_budha}. \\

To study the ground state properties, we analyze the one-body density and the occupation of the natural orbitals, derived from the reduced one-body density matrix (RDM). The diagonal of the RDM gives the spatial density, while its eigenvalues reveal whether the system is condensed (with a single macroscopic eigenvalue) or fragmented (i.e., having multiple significant eigenvalues)~\cite{rhombik_acc,Penrose,anal_2020_dw}.  The main text highlights the key features of the fragmentation, while the supplementary material provides further details and the depletion dynamics.

To study the transport properties, we compute the position expectation values in both the $x$- and $y$-directions. 
The position expectation value in the x-direction is given by $\langle x \rangle = \frac{1}{N}\langle \Psi(t) \vert \hat{X} \vert \Psi(t) \rangle$, where $\hat{X} = \sum_{j=1}^N \hat{x}_j$. 
We compute $\langle x \rangle$ using the mean-field and many-body wave-function and define them as $\langle x \rangle_{MF}$ and $\langle x \rangle_{MB}$,  respectively. The corresponding expectation value in the $y$-direction, $\langle y \rangle$, is obtained analogously. These expectation values provide a direct measure of the collective displacement of particles over time.

The variance of position operators serve as a crucial measure in  the quantum transport, as it quantifies both the spatial spread and the fluctuations in the system. It also helps distinguish between coherent and incoherent transport and provides insights into the directional anisotropy during the transport process, particularly identifying when transport becomes suppressed or frozen in certain directions~\cite{anal_2020_dw}. The variance per particle of the position operator $\hat{X}$, denoted by $\frac{1}{N} \Delta_{\hat{X}}^2$, is defined as
\begin{eqnarray}
\frac{1}{N}\Delta_{\hat{X}}^2 = \frac{1}{N}  \langle \Psi(t) \vert \hat{X}^2 \vert \Psi(t) \rangle - \frac{1}{N}\langle \Psi(t) \vert \hat{X} \vert \Psi(t) \rangle^2 .
\end{eqnarray}
As previously discussed, the second term contains contributions only from one-body operators, whereas $\langle \hat{X}^2 \rangle$ includes both one-body and two-body contributions: $\hat{X}^2 = \sum_{j=1}^N \hat{x}_j^2 + \sum_{k>j=1}^N 2\hat{x}_j\hat{x}_k$.
This two-body term makes the variance particularly sensitive to correlations in the system. Therefore, it plays an imperative role in the many-body analysis. For more details, see Ref.~\cite{variance_ofir2015,variance_ofir2019_symmetry,variance_ofir2019}.
Using the same framework, we can extend our analysis to quantify the position variance along the y-direction, as well as the momentum and angular momentum variances. While the main text focuses on the position variances in both x- and y-directions, a complementary set of analyses---covering momentum and angular momentum variances---is presented in the supplementary material.

\section{Results and discussion}\label{results}
We divide the results section into three parts. First, we examine how bending affects the ground-state properties of bosons confined in the two-dimensional bent potential. Specifically, we examine the ground-state density distribution and interaction-induced correlations across various bending geometries. 
In the second part, we study how bosons behave when transported through the bent potential. We use both mean-field and many-body methods to examine how different properties change as the bosons navigate through the bend. 
The final part shows how transport can be tuned solely by modifying different parameters of the bend geometry. Our analysis primarily relies on position expectation values and position variances. Throughout this study, we consider a system of $N=10$ bosons with the mean-field interaction parameter $\Lambda=0.1$. It is worth noting that in mean-field calculations,  $\Lambda$ is the defining parameter, while the actual particle number $N$ can, in principle, be arbitrarily large.

\subsection{Statics}\label{static}
We begin by studying the ground-state behaviour of the bosons confined in the bent potential. 
We compute the ground state of the Hamiltonian in Eq.~(\ref{hamiltonian}) by propagating $\hat{H} \Psi =i\frac{\partial \Psi}{\partial t}$ in imaginary time. 
The parameters are set as $L=0.5$, $D=5.0$, $a=7.0$ and $V_0 = 10.0$, while $B$ is a variable. These parameter choices generate a smooth bend in both directions which prevents unwanted wave reflections at the sharp edges, while still being pronounced enough to reveal curvature-induced effects. The choices also provides ample space along the y-direction. For a fixed $D$, increasing $B$ results in a more pronounced bending of the potential. We investigate how this bending influences the ground state properties of the system.

\begin{figure}
    \centering
    \includegraphics[width=0.75\textwidth, angle =-0 ]{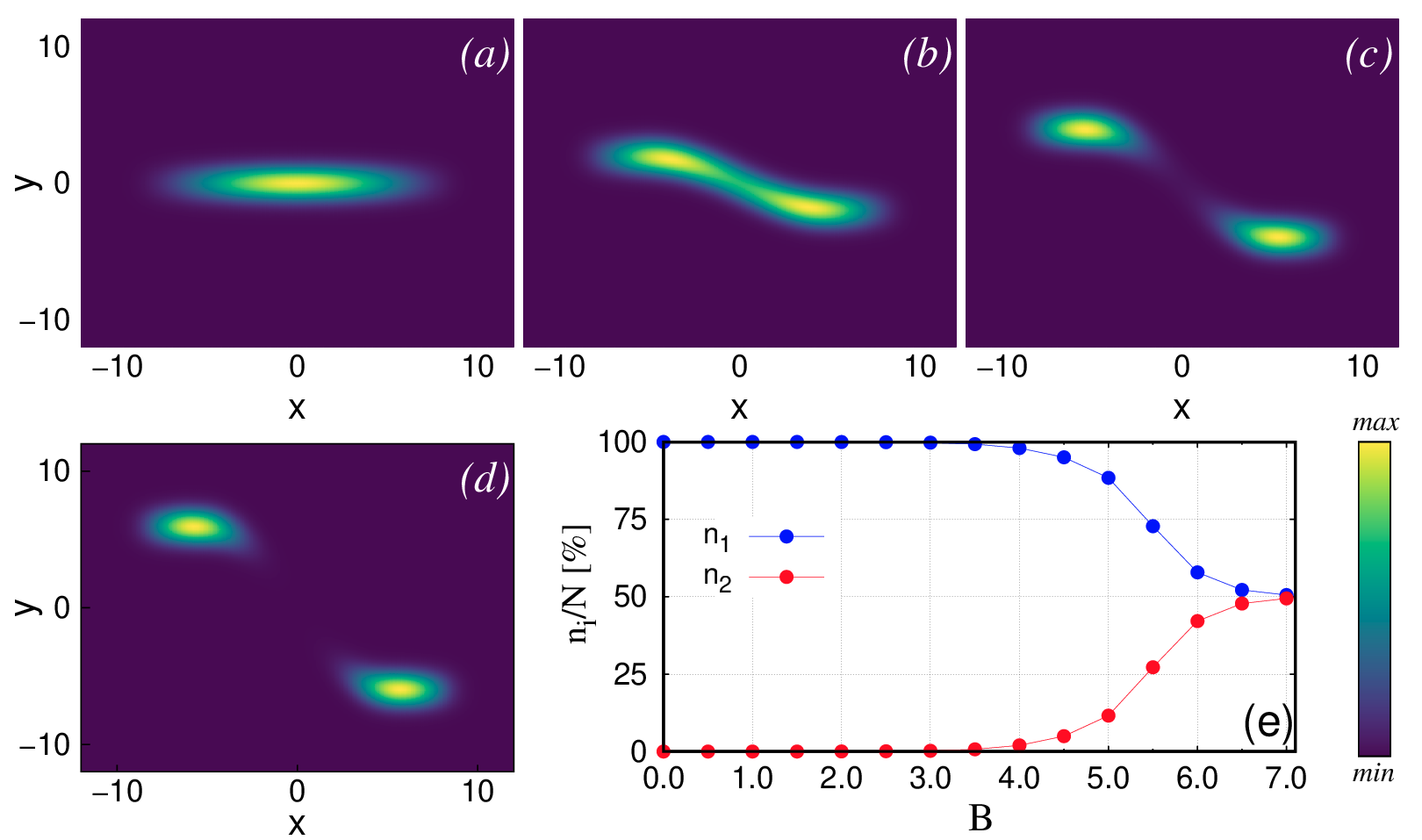}
    \caption{($a-d$) One-body density of the ground state in the bent potential for different values of the half-width $B$. As $B$ increases, the density gradually splits into two distinct, localized lobes. The effect becomes more pronounced at larger $B$: (a) $B=0$, (b) $B=2.0$, (c) $B=4.0$, and (d) $B=6.0$.  While the results shown are from the many-body calculations, the mean-field densies exhibit similar structures. ($e$) Occupation in the first two natural orbitals as a function of $B$, showing loss of coherence with increase in the width of the bent. For sufficiently large $B$, the system exhibits 50\% fragmentation. See text for further discussion. All quantities are dimensionless.}
    \label{fig1}
\end{figure}
The calculated one-body densities are shown in Fig.~\ref{fig1}(a–d). As the half-width of the bent increases, the density evolves from being fully delocalized at $B=0.0$ to becoming localized at two distinct positions. This behavior resembles that of a conventional double-well potential, where increasing the barrier height leads to density splitting~\cite{PhysRevA.59.3868}. Similar splitting has also been reported for bosons under rotation trapped in rotationally asymmetric anharmonic traps~\cite{sunayana_scirep,rhombik_scirep}. 
But, in this study, we focus specifically on understanding how the curvature of the potential influences the localization behavior of the bosons. To illustrate this effect, we mainly vary a single geometric parameter---the half-width $B$---which determines the extent of the bend. However, it is important to note that other parameters, such as the sharpness of the bent ($D$), the width of the middle flat region ($2L$) and the smoothness of the transverse confinement ($a$), also provide additional means to control localization. This tunability enhances the versatility of the setup, allowing not only for the emulation of traditional double-well potentials, but also for the design and exploration of more intricate multi-well configurations by incorporating multiple bends into the potential landscape.

The densities shown here are obtained from the many-body method, and the corresponding mean-field results display a similar structure. Since fragmentation is inherently a many-body effect, its variation with the half-width of the bent potential is illustrated in Fig.~\ref{fig1}($e$). The figure displays the occupations of the first two natural orbitals. As the half-width $B$ of the bent increases, the system transitions from a condensed to a fragmented state, evidenced by a decrease in $n_1$ and a corresponding increase in $n_2$. This is an analogue to the double well setup where increase in the barrier height can cause fragmentation at high barrier height~\cite{PhysRevA.59.3868, PhysRevA.96.053627}. In Fig.~\ref{fig1}(e), only the first two orbital occupations are shown, as they are the most significantly populated. The occupations of the remaining orbitals are negligible, of the order of $10^{-5}$.

Thus, as the bending increases, the density develops two prominent lobes, leading to fragmentation of the many-body wave function. Despite this structural change, the expectation values for both position and momentum remain zero along the x- and y-directions. This behavior reflects the underlying $C_2$ symmetry of the potential despite the emerging localization and fragmentation phenomena. The ground-state energy is found to increase as $B$ increases. It is also observed that (not shown), as the system becomes fragmented,  the discrepancy between the mean-field and many-body energies grows with increase in the width of the bent, with the many-body energy always below the mean-field energy, of course. 

\begin{figure}
    \centering
    \includegraphics[width=0.95\textwidth, angle =-0 ]{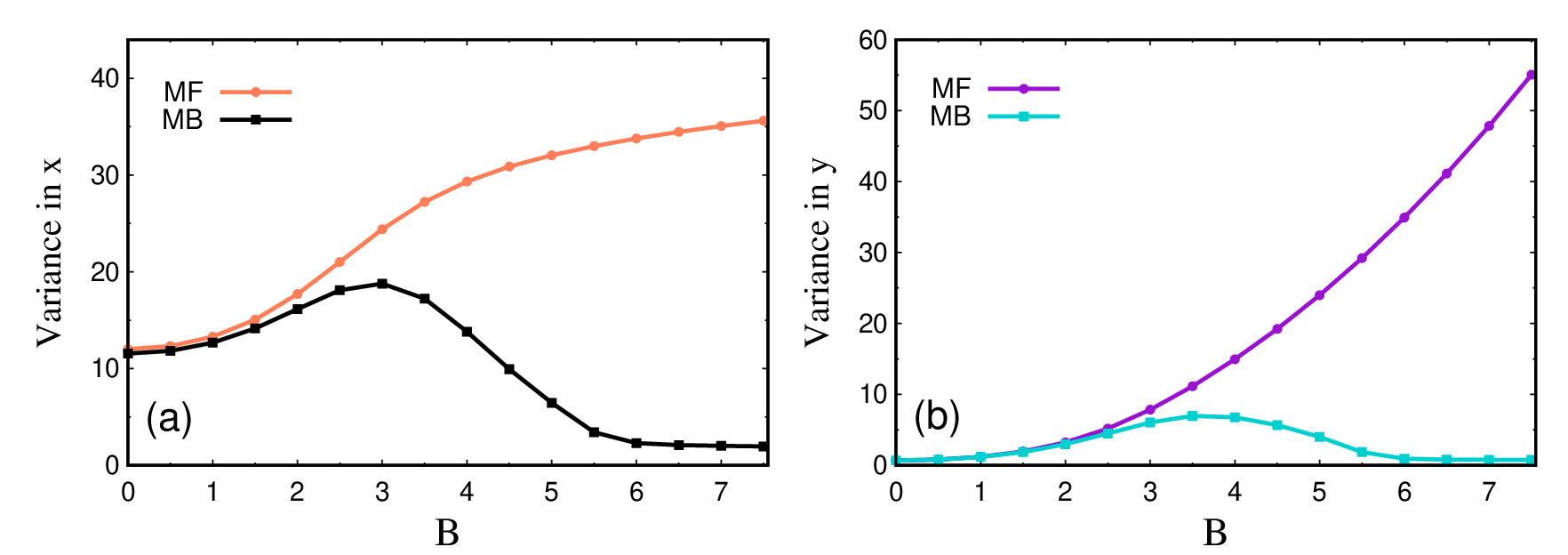}
    \caption{(a) Position variance in the x-direction, $\frac{1}{N} \Delta_{\hat{X}}^2$, and (b) position variance in the y-direction, $\frac{1}{N} \Delta_{\hat{Y}}^2$, as functions of the half-width of the bent. As $B$ increases, the mean-field variances grow, while the many-body variances initially rise, then begin to decline beyond a certain threshold of $B$. This contrasting behaviour reflects the onset of fragmentation within the system. See the text for more details. All quantities are dimensionless.}
    \label{fig2}
\end{figure}

To deepen our understanding, we analyze the position variances of the system as functions of the half-width, at both the mean-field and many-body levels. The position variances along the x- and y-directions are presented in Fig.~\ref{fig2}, serving as a sensitive probe of quantum fluctuations and correlations. The many-particle position variance in the x-direction, $\frac{1}{N}\Delta_{\hat{X}}^2$, is shown in Fig.~\ref{fig2}(a). When there is no bending ($B=0$), the density cloud remains symmetrically delocalized around $x=0$.  As the bending increases, the density splits into two localized regions, and the overall spatial spreading of the cloud expands (see Fig.~\ref{fig1}). Consequently, both the mean-field and many-body x-variances increase with $B$. However, the confinement in the x-direction restricts the bosons from dispersing further, which slows down the growth of the mean-field x-variance at larger $B$. For small widths, the mean-field and many-body variances remain nearly identical, signify weak correlations. Up to approximately $B \sim 3.0$, both variances grow, but the mean-field result exhibits a steeper rise. Beyond this threshold, their behaviours diverge: the mean-field variance continues to increase, whereas the many-body variance begins to decrease.

A similar pattern is observed for the variance in the y-direction, $\frac{1}{N}\Delta_{\hat{Y}}^2$, shown in Fig.~\ref{fig2}(b). As $B$ increases, the density expands in the y-direction, leading to a steady, monotonic increase in the mean-field variance. In contrast, the many-body variance again shows a non-monotonic trend: it increases up to $B \sim 3.0$, after which it gradually decreases. This drop in many-body variance in both spatial directions indicates the onset of significant many-body effects. This noticeable reduction in the many-body variance closely aligns with the onset of fragmentation [see Fig.~\ref{fig1}(e)]. 
In the typical $2D$ double-well systems where the double well is generally aligned along one axis, the difference between mean-field and many-body variance appears only along that axis---the other direction remains unchanged~\cite{variance_ofir2019}. However, in our case, the double-well-like structure develops because of the bent and is not aligned with either the x- or y-axis, leading to differences between mean-field and many-body results in both directions. 
Notably, the similar behaviour of the variance in both spatial directions also suggests that excitations---and hence correlations---are manifest along both x- and y-direction. This observation raises an interesting question: how do these excitations influence the transport properties of the system?

The analysis of the density, occupations in the natural orbitals, and the position variances provides a comprehensive understanding of the ground-state characteristics of the bent potential. These findings highlight the natural emergence of a double-well-mimicking structure within the bent potential. Guided by this perspective, the next section is devoted to investigating the transport dynamics of bosons across the bent.

\subsection{Transport} \label{transport}
The ground-state calculations reveal that the bent in the two-dimensional channel potential acts as an effective barrier, leading to spatial localization of the density into two distinct regions as the width of the bent increases. To study the transport, we initially confine the bosons to the left side of the bent potential, then release them and monitor their dynamics. Mathematically, this is achieved by preparing the bosons on the left using a longitudinal sinusoidal barrier at $x=0$. The system is then evolved in the full bent potential as the barrier is abruptly shifted from $x=0$ to $ x=+8$. We observe pronounced back-and-forth motion of bosons between the upper-left and lower-right domains. 

To characterize transport dynamics, we employ the time-dependent position expectation values along both in x- and y-directions. Since the bosons primarily undergo oscillations between the two spatial domains, the survival probability serves as a natural measure for probing tunneling phenomena~\cite{anal_2022_dw,rhombik_acc,rhombik_epjd}. However, while survival probability effectively captures tunneling events, it inherently lacks spatial resolution and provides little insight into the redistribution or evolution of density across the potential landscape. In contrast, the position expectation values offer a direct indicator of spatial displacement, making them particularly well-suited for analyzing the transport through the bent potential. For completeness, a detailed analysis of the survival probability is provided in the supplementary material.

\begin{figure}
    \centering
    \includegraphics[width=0.95\textwidth, angle =-0 ]{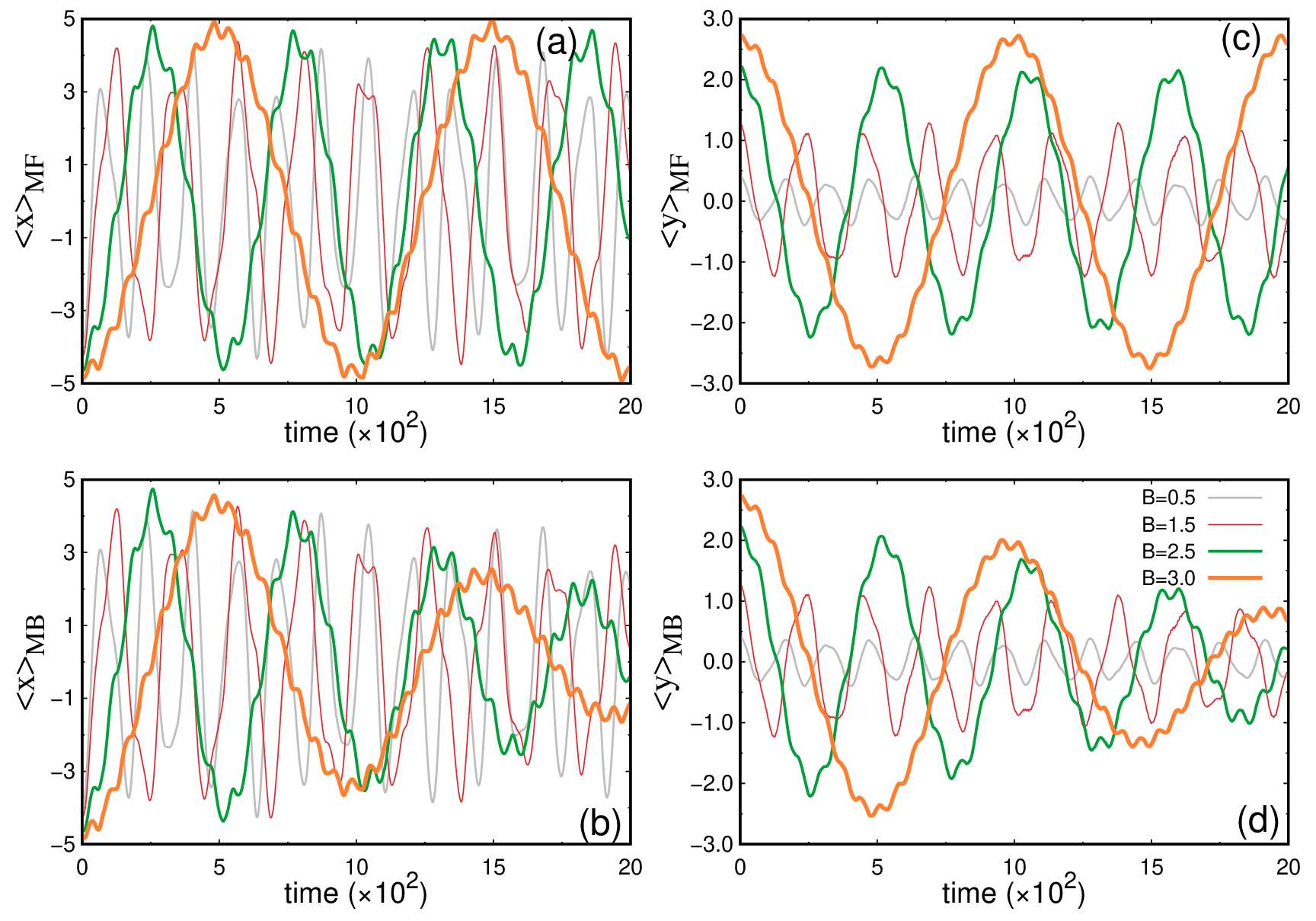}
    \caption{Position expectation values along the x- and y-directions are presented. Panels ($a$) and ($b$) display the mean-field and many-body $\langle x \rangle$, respectively, while panels ($c$) and ($d$) show the corresponding values for $\langle y \rangle$. The transport dynamics are investigated by analyzing the particle flow through the bent structure. The results reveal back-and-forth motion of the bosons between the upper-left and lower-right sides of the bent. As the width of the bent increases, the oscillations become more regular and the oscillations' period increases. Additionally, the many-body expectation values exhibit a reduction in oscillations' amplitude due to the onset of fragmentation. All quantities are dimensionless. Further details are discussed in the text.}
    \label{fig3}
\end{figure}

Figure~\ref{fig3} presents the time evolution of the position expectation values in the x- and y-directions, computed using both mean-field and many-body densities, for different values of the half-width $B$. All other geometric parameters are kept fixed: the sharpness of the bent ($D=5.0$), the length of the middle flat segment ($2L=1.0$), and the transverse confinement strength ($a=7.0$). The mean-field expectation value $\langle x \rangle_{MF}$ is shown in Fig.~\ref{fig3}(a) and the many-body calculation $\langle x \rangle_{MB}$ is presented in Fig.~\ref{fig3}(b) for four different half-widths of the bent ($B=0.5,1.5$, $2.5$ and $3.0$). 
Initially at $t=0$, bosons are trapped in the upper-left side of the trap.  
Following the quench, particles move through the bent and gradually accumulate on the opposite side. They then tunnel back through the bent, accumulating again on the original side. This oscillatory transport dynamics continues throughout the entire observed time window. The influence of the width of the potential on tunnelling is evident: as $B$ increases, the tunnelling time becomes longer, and the oscillatory behaviour of $\langle x \rangle$ becomes smoother and more regular. 

In the many-body results, a notable reduction in oscillations' amplitude is observed across all $B$, with the effect becoming more pronounced for larger $B$. This is attribute to the gradual loss of coherence in the system---a phenomenon well-documented in studies of tunnelling dynamics~\cite{Universality_of_fragmentation,rhombik_acc,anal_2022_dw}. 
The mean-field expectation value $\langle y\rangle_{MF}$ is shown in Fig.~\ref{fig3}(c) while the corresponding many-body calculation $\langle y \rangle_{MB}$ is presented in Fig.~\ref{fig3}(d), both evaluated for the same set of $B$ as previously discussed. 
We can assert that since the double-well-like structure induced by the bend is not aligned with any specific axis, oscillations appear in both $\langle x \rangle$ and $\langle y \rangle$. The expectation value of the angular momentum, $\langle l_z \rangle$, also exhibits small-amplitude oscillations, as shown in the supplementary material. This behavior highlights the inherently two-dimensional nature of the double-well-like structure, where the angular momentum remains nonzero during the dynamics. 
A key distinction from the $\langle x \rangle$ dynamics is that the oscillation amplitude varies with $B$. This behavior is natural because a larger $B$ corresponds to a wider bent, allowing the bosons to explore more space along the y-direction. The many-body dynamics exhibit a gradual decrease in oscillations' amplitude over time---in analogy to the behavior observed along the x-direction---reflecting the emergence of  many-body effects. We analyze how these effects develop and evolve following the quench below, with further details in the supplementary material. 

We have also observed a beating pattern in the oscillatory motion (see Fig.~\ref{fig3}), which indicates the presence of multiple frequencies during the tunneling process. This phenomenon arises from the admixture between different energy modes contributing to the transport dynamics. The detailed origin and role of other frequency components are thoroughly analyzed in Appendix~\ref{apndx}. Additionally, a full video of the density dynamics is provided in the supplementary material.  

\begin{figure}
    \centering
    \includegraphics[width=0.75\textwidth, angle =-0 ]{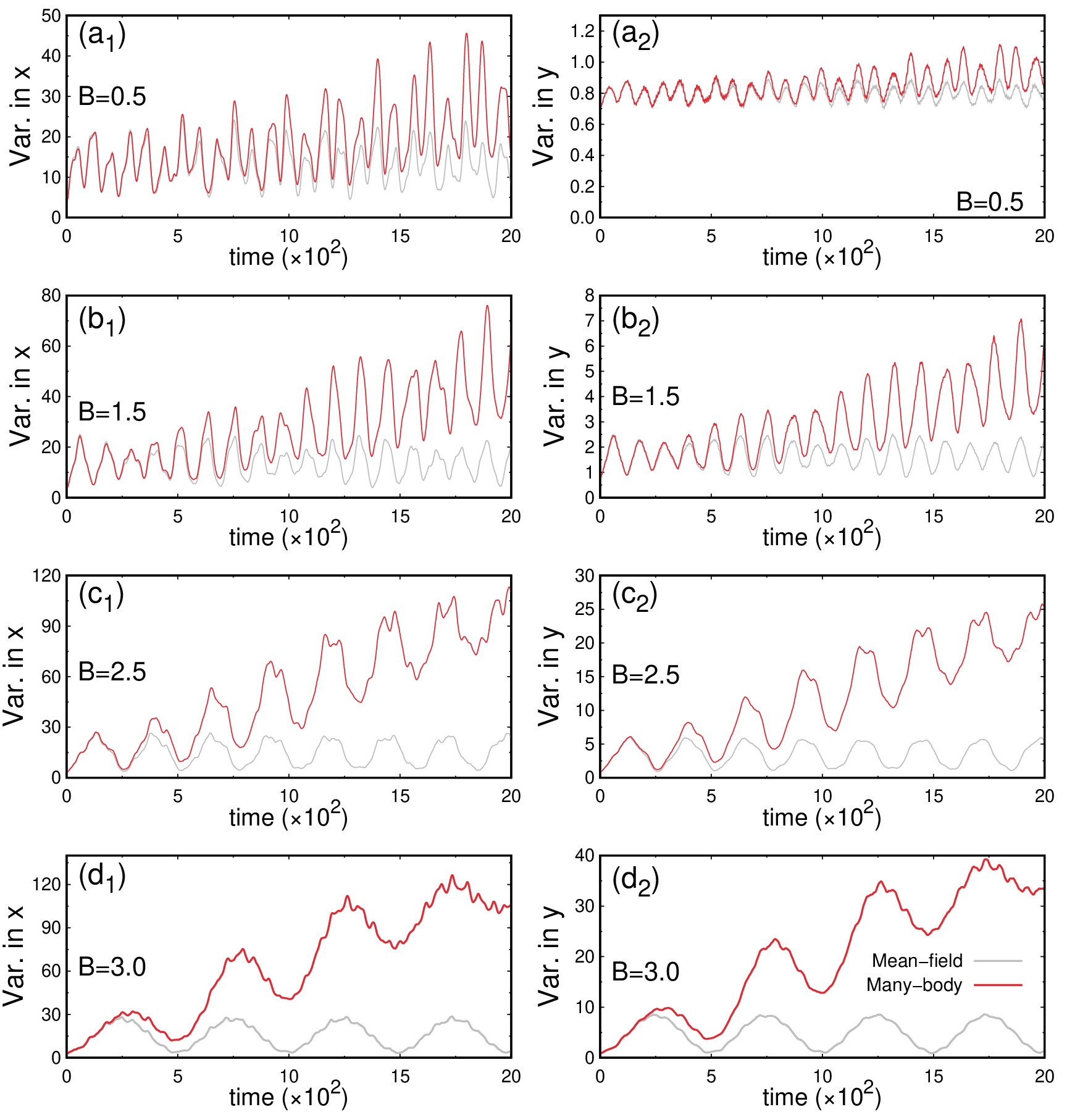}
    \caption{Time evolution of the variances in the x-direction (panels $a_1, b_1, c_1,d_1$) and y-direction (panels $a_2, b_2, c_2, d_2$) for four values of the half-width of the bent $B$: ($a_1,a_2$) $B=0.5$, ($b_1,b_2$) $B=1.5$, ($c_1,c_2$) $B=2.5$, and ($d_1,d_2$) $B=3.0$. In all cases, the mean-field variances exhibit a smooth, bound-like oscillatory dynamics, while the many-body variances increase with time in an oscillatory manner.  The growing difference between the mean-field and many-body variances over time reflects the emergence of fragmentation in the system and its deviation from mean-field behavior. Note that the vertical scales differ in each panel to better visualize the differences between mean-field and many-body variances. See text for further details. All quantities are dimensionless. }
    \label{fig4}
\end{figure}

Figure~\ref{fig4} provides another instructive insight into the interplay between the width of the bent and the many-body quantum dynamics in the bent potential. The panels show the time-dependent variances in the longitudinal and transverse directions for four different values of $B$, revealing a systematic transition in the transport behavior as the geometry is tuned. We now examine the transport behavior for each bent configuration individually. In the smallest bent ($B=0.5$, panels $a_1$ and $a_2$), both mean-field and many-body variances evolve in close synchrony, displaying low-amplitude oscillations and minimal deviation from one another. This tight agreement manifests weak fragmentation and validates the applicability of the mean-field description, at least in short timescales. A detailed quantitative analysis of the depletion dynamics is provided in the supplementary material. As the bent broadens to $B=1.5$, a pronounced difference between the many-body and mean-field dynamics becomes evident. The many-body variances start to increase more rapidly than the previous case. This reflects the onset of fragmentation during passing through the bent. For $B=2.5$ and $B=3.0$, the many-body position variances exhibit significant differences along both the x- and y-directions—clear signatures of strong correlations and fragmentation in the quantum dynamics. In contrast, the mean-field variances remain smooth and suppressed, highlighting the limitations of mean-field theory in capturing the complex behavior of the system in this regime.

Thus, the time evolution of the position expectation values and the variances provide a clear, quantitative evidence that the geometry of the potential directly controls the degree of many-body effects during the transport. Wider bents promote fragmentation which are absent or suppressed in the narrow bent regime. 
This behavior shows that the geometry of the system---especially the half-width of the bent---can be used as a  tool to control quantum coherence and fragmentation. In the next section, we show that it is not just the half width of the bent that matters; other geometric parameters can also be tuned to influence transport. This makes the bent potential a versatile platform for controlling quantum transport through simple geometric adjustments.

Analysis of the expectation values and the spatial variances reveals that transport in the bent potential exhibits clear signatures of back-and-forth motion between the two sides, closely resembling the dynamics observed in a conventional double-well system. The  difference is that the emergence of the double well due to the bent is not aligned with a specific axis, which leads to oscillations in the position expectation values along both directions. Side by side, some angular momentum dynamics sets in during the transport. 

For weakly interacting bosons trapped in a double well potential, the characteristic tunneling period is governed by the energy gap between the lowest two eigenstates, $T=\frac{2\pi}{E_2 - E_1}$, where $E_1$ is the ground state energy and $E_2$ is the first excited state energy. 
To assess whether the two-level description remains valid for the dynamics in the bent potential, we computed a few lower excited-state energies using exact diagonalization. The corresponding energy gaps yield tunneling timescales that closely match the numerical simulations. Additional frequencies appear in the time evolution of the position expectation values, suggesting contributions from higher excited modes. We examine this behavior in detail in Appendix~\ref{apndx}. Importantly, the energy gap can be precisely controlled by varying the geometric parameters of the potential. Thus the oscillation period can be tuned to any desired value by changing the potential parameters discussed in Sec.~\ref{bent_pot}.

\begin{figure}
    \centering
    \includegraphics[width=0.95\textwidth, angle =-0 ]{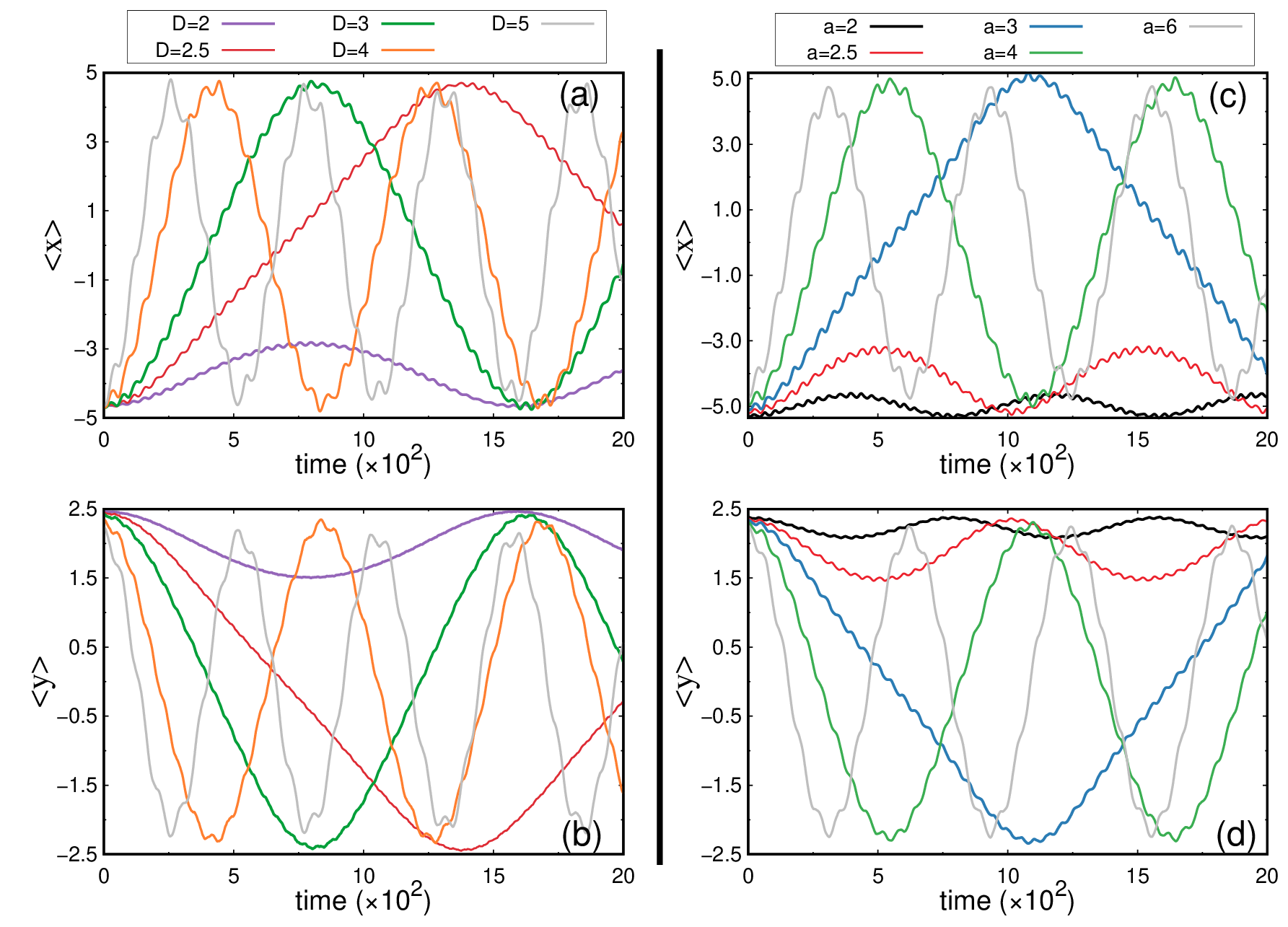}
    \caption{Position expectation values along the x- and y-directions are shown to illustrate the controllability of the transport in the bent potential. Panels (a) and ($b$) explore the effect of varying the sharpness of the bent ($D$), while panels ($c$) and ($d$) examine the role of transverse confinement ($a$). The dynamics range from self-trapping to complete oscillations, determined by the parameters. In the full tunnelling regime, $D$ and $a$ serve as tuning knobs for the oscillation period. The calculations are performed using the mean-field method. The corresponding many-body results exhibit dynamics on the same timescale, but the oscillations' amplitudes decay over time.  See the text for further details. All quantities are dimensionless. }
    \label{fig6}
\end{figure}

\subsection{Curvature-driven control of quantum transport}

The previous sections examined both the static and dynamic properties of the bent potential, revealing that the dynamics effectively behaves like a double-well system. The tunneling period is highly tunable through adjustments to the width of the bent. In this section, we move beyond this limit to examine how the transport can be actively engineered by leveraging the full geometric flexibility of the bent potential.

Unlike conventional double-well potentials---where tunneling is primarily dictated by the height and width of the central barrier---the bent potential provides a richer landscape with multiple independent control knobs. The geometry of the bent potential is mainly governed by four principal parameters: the half-width $B$, the sharpness of the bent $D$, the middle flat region $2L$, and the transverse confinement $a$, see Fig.~\ref{fig0}. Throughout our analysis, we fix $L=0.5$. While the earlier section focused on the influence of $B$, we now investigate how tuning $D$ and $a$ enables precise modulation of tunneling behavior and overall transport properties.

Figures~\ref{fig6}(a,b) show the expectation values of $\langle x \rangle$ and $\langle y \rangle$ for different sharpness of the bent $D$, while keeping the other parameters fixed at $a=7.0$ and $B=2.5$. For a very sharp bend (small $D$), the system remains fully self-trapped, i.e., tunneling between the two sides of the bent is strongly suppressed. As the bend becomes smoother (moderate $D$), the bosons begin to partially tunnel through. when $D=$ increases further to $2.5$, the bosons tunnel completely. Making the bend even smoother (further increasing $D$) reduces the tunneling time period. Thus, the sharpness of the bent serves as an additional control parameter, allowing transitions from a self-trapped regime to complete tunneling and offering a way to tune the tunneling timescale.

The tunneling process is influenced not only by the width and sharpness of the bend but also significantly by the transverse confinement. In Fig.~\ref{fig6}(c,d), we illustrate how variations in the transverse confinement can drive the system from self-trapping to controlled tunneling with a well-defined time period by looking at the mean-field $\langle x \rangle$ and $\langle y \rangle$. The other parameters are fixed at sharpness $D=5.0$ and half-width $B=2.5$. The figures show that for strong transverse confinement (i.e., small $a$), the system exhibits self-trapping. As the confinement becomes less stiff (moderate $a$), tunneling begins to occur partially. At $a=3.0$, complete tunneling is observed. Increasing $a$ further leads to a decrease in the tunneling time period.

While the results presented here are based on mean-field calculations, corresponding many-body simulations (not shown) reveal that the tunnelling period remains similar; however, the amplitude of oscillations  reduces with time. As discussed in the previous section (Sec.~\ref{transport}), this suppression arises from the build-up of condensate depletion, which reflects the onset of many-body correlations beyond the mean-field description. The evolution of the depletion and its impact on the transport dynamics are analyzed in details in the supplementary material.

In summary, the bent potential offers a controllable platform for engineering tunnelling dynamics. By adjusting the sharpness of the bent and the transverse confinement, one can drive the system from localized to fully tunnelling regimes. The sensitivity of the transport to these geometric parameters along with the emergence of many-body effects highlight the bent potential as a versatile tool for studying correlated quantum transport and designing quantum devices with tunable coherence properties.

\section{Conclusions}\label{conclusion}

Quantum transport in ultracold bosonic systems is a subject of significant interest. In this work, we explore the transport properties of weakly interacting bosons in a two-dimensional bent potential, uncovering rich tunneling dynamics governed by the confinement geometry. Our study examines both the ground state and the time-dependent transport behavior, analyzed using the many-body multi-configurational time-dependent Hartree method for bosons. 

The ground-state characteristics of the bosons confined in the bent potential exhibit a pronounced sensitivity to the geometric deformation of the potential. As the half-width of the bent increases, the density evolves from being delocalized across the entire bent to becoming spatially separated, and thereby forming two distinct lobes. This spatial restructuring is accompanied by the loss of coherence. 

In the dynamical regime, the transport behavior is analyzed both from the mean-field and many-body perspectives. First, we studied how the quantum transport emerges as the half-width of the bent is increased. By tracking the time evolution of the position expectation values, we observed that the tunnelling time period highly depends on the bent structure. In the many-body analysis, the oscillations' amplitude decreases over time due to condensate depletion, highlighting the influence of beyond-mean-field effects. Position variance analysis further revealed the loss of correlations during the evolution. Additionally, the oscillation frequencies derived from the exact diagonalization energy levels shows close agreement with the numerical results. 
We also probed the sensitivity of the bent potential to additional confinement parameters, showing that both the sharpness of the bent and the transverse confinement strongly affect the tunnelling dynamics. Varying these parameters enables fine control over the amplitude and timescale of transport. Although the bent potential is structurally different from a double well, it exhibits comparable dynamical behavior. Moreover, the enhanced geometric tunability of the bent potential establishes it as a versatile platform for controllable quantum transport.

All in all, the bent potential mimics double-well–like dynamics with exceptional geometric tunability, providing a powerful platform for controlling quantum transport, accessing strongly correlated regimes, and probing coherence and fragmentation. Looking forward, extending this setup to include multiple bends could create a series of transport barriers, effectively forming a synthetic optical lattice within the curved geometries. Additionally, incorporating long-range interactions is expected to produce nontrivial effects in combination with the curved potential landscape. These prospects open new avenues for exploring lattice-like physics in engineered curved potentials, offering promising directions for both theoretical and experimental research.

\section*{Acknowledgements}
This work is supported by the Israel Science Foundation (ISF) grant no. 1516/19. Computation time at the High-Performance Computing Center Stuttgart (HLRS) and High Performance Computing system Hive of the Faculty of Natural Sciences at University of Haifa are gratefully acknowledged.

\appendix
\section{Exact diagonalization analysis}\label{apndx}
In this section, we analyze the spectrum of a single boson confined in the bent potential using exact diagonalization (ED). The ground and excited state energies provide insight into the structure of the spectrum as a function of the half-width $B$, and helps us to understand the behaviour of the tunnelling dynamics observed in Sec.~\ref{transport}.

Table~\ref{tab1} lists the energies of the ground and a few low-lying excited states for four different values of $B$. The other parameters are set at $L=0.5$, $D=5.0$ and $a=7.0$. It is observed that, as the width of the bent  increases, the energy gaps between the first two levels ($E_1$ and $E_2$) and between the next two levels ($E_3$ and $E_4$) become smaller. This trend is analogue to the behavior in a double-well potential, where increasing the barrier height reduces the splittings $E_2 - E_1$ and $E_4 - E_3$, leading to the formation of nearly degenerate doublets that eventually evolve into energy bands. 
Ground-state energies per particle calculated from MCTDHB closely match ED results, with minor deviations due to the finite interaction parameter $\Lambda=0.1$ used  in the MCTDHB calculation. Of course, in the non-interacting limit, $\Lambda \rightarrow 0$, the many-body description converges to the exact single-boson result. 

\begin{table}[htbp]
    \centering
    \begin{tabular}{||c|c|c|c|c||}
        \hline \hline
         & $B=0.5$ & $B=1.5$ & $B=2.5$ & $B=3.0$ \\
        \hline
        $E_1$ & 0.3126954673319  & 0.3269747042284 & 0.3473962003652 & 0.3562600501999   \\ \hline
        $E_2$ & 0.3519929098011 & 0.3541570050852 & 0.3591726151522 & 0.3625937085079 \\ \hline
        $E_3$ & 0.4205157045709 & 0.4268299856663 & 0.4459900888471 & 0.4627963861842 \\ \hline
        $E_4$ & 0.5145407802742 & 0.5146742968526 & 0.5192992004093 & 0.5246286782089 \\ \hline
        \hline
    \end{tabular}
    \caption{Low-lying energy levels of a single particle in a bent potential well, calculated using exact diagonalization. Results are shown for four different half-widths of the bent.}
    \label{tab1} 
\end{table}

The decrease in the energy gap between the ground and first excited states as the half-width of the bent increases indicates a reduction in the effective coupling between the two sides of the bent potential. This corresponds to the particle becoming more localized within individual sides, rather than being delocalized across the entire bend. Such behavior is consistent with the emergence of well-separated density lobes observed in the ground state, shown in Sec.~\ref{static}. 
The spectral signatures indicate that the bent geometry effectively behaves like a double-well potential, with $B$ serving as an effective barrier height. This naturally directs our attention to the tunneling dynamics, which in the weakly interacting regime are likely governed by the low-lying energy differences.\\

\begin{table}[htbp]
    \centering
    \begin{tabular}{||c|c|c||}
        \hline \hline
         & \makecell{Frequencies obtained \\ from the numerics} & \makecell{Corresponding ED matching \\ $f=\frac{(E_i -E_j)}{2 \pi}$}\\
        \hline
        $B=0.5$ & \makecell{6.245 $\times 10^{-3}$ \\ 1.099 $\times 10^{-2}$ } 
        & \makecell{6.254$\times 10^{-3}$ ($E_2 - E_1$) \\ 1.090$\times 10^{-2}$ ($E_3 - E_2$) }
        \\ \hline
        $B=1.5$ & \makecell{4.330 $\times 10^{-3}$  \\ 1.157 $\times 10^{-2}$ } 
        & \makecell{ 4.326$\times 10^{-3}$ ($E_2 - E_1$) \\ 1.156$\times 10^{-2}$ ($E_3 - E_2$)  }
        \\ \hline
        $B=2.5$ & \makecell{1.903 $\times 10^{-3}$ \\ 1.370 $\times 10^{-2}$} & \makecell{ 1.874$\times 10^{-3}$ ($E_2 - E_1$)  \\ 1.381$\times 10^{-2}$ ($E_3 - E_2$)}  \\ \hline
        $B=3.0$ & \makecell{0.999 $\times 10^{-3}$ \\ 1.570 $\times 10^{-2}$} & \makecell{ 1.008$\times 10^{-3}$ ($E_2 - E_1$)  \\ 1.594$\times 10^{-2}$ ($E_3 - E_2$)}  \\ \hline
        \hline
    \end{tabular}
    \caption{Comparison between the dominant oscillation frequencies obtained from the Fourier transform of the numerically calculated position expectation values and the analytical frequencies determined from the energy differences $E_i - E_j$ obtained using exact diagonalization. All quantities shown are dimensionless.}
    \label{tab2} 
\end{table}

To establish a connection between the static spectral properties and the  dynamics, we performed a Fourier analysis of the numerically computed time evolution of both $\langle x \rangle$ and $\langle y \rangle$. 
The Fourier analysis is performed over a time interval which is ten times longer than that presented in the results section, to accurately resolve the oscillation frequencies ($\Delta f \approx 5 \times 10^{-5}$). The oscillations' period depends on the parameter $B$, with the longest period observed at $B=3.0$. To achieve adequate frequency resolution, the simulation for $B=3.0$ is performed over at least $20$ oscillation cycles, while proportionally more cycles are included for smaller $B$ values. 
This analysis reveals that, across all cases, the motion is dominated by two main oscillation frequencies with exhibiting identical frequency components.
These dominant frequencies are compared with those obtained from the energy differences $E_i - E_j$  calculated using the ED method. Table~\ref{tab2} summarizes this comparison, with the left column showing the numerically extracted frequencies from the dynamics and the right column shows the corresponding frequencies obtained from the energy gaps computed using ED. We found mainly two frequencies during the dynamics. The close match confirms that a few low-lying eigenstates contribute to the transoprt dynamics~\cite{PhysRevA.78.013621, PhysRevA.87.043626}. 

The tunneling dynamics is mainly governed by the frequency $f_1=\frac{(E_2 -E_1)}{2 \pi}$, corresponding to the complete oscillations between the two sides of the potential. A weaker secondary component at $f_2=\frac{(E_3 -E_2)}{2 \pi}$ arises from a small admixture of the second and third eigenstates, producing small beating in the position expectation values. This secondary contribution becomes even weaker as the half-width of the bent is increased, indicates that broader geometries better confine the dynamics to the lowest two modes and suppress coupling to higher states~\cite{PhysRevA.82.063626}.\\

\begin{figure}
    \centering
    \includegraphics[width=0.8\textwidth, angle =-0 ]{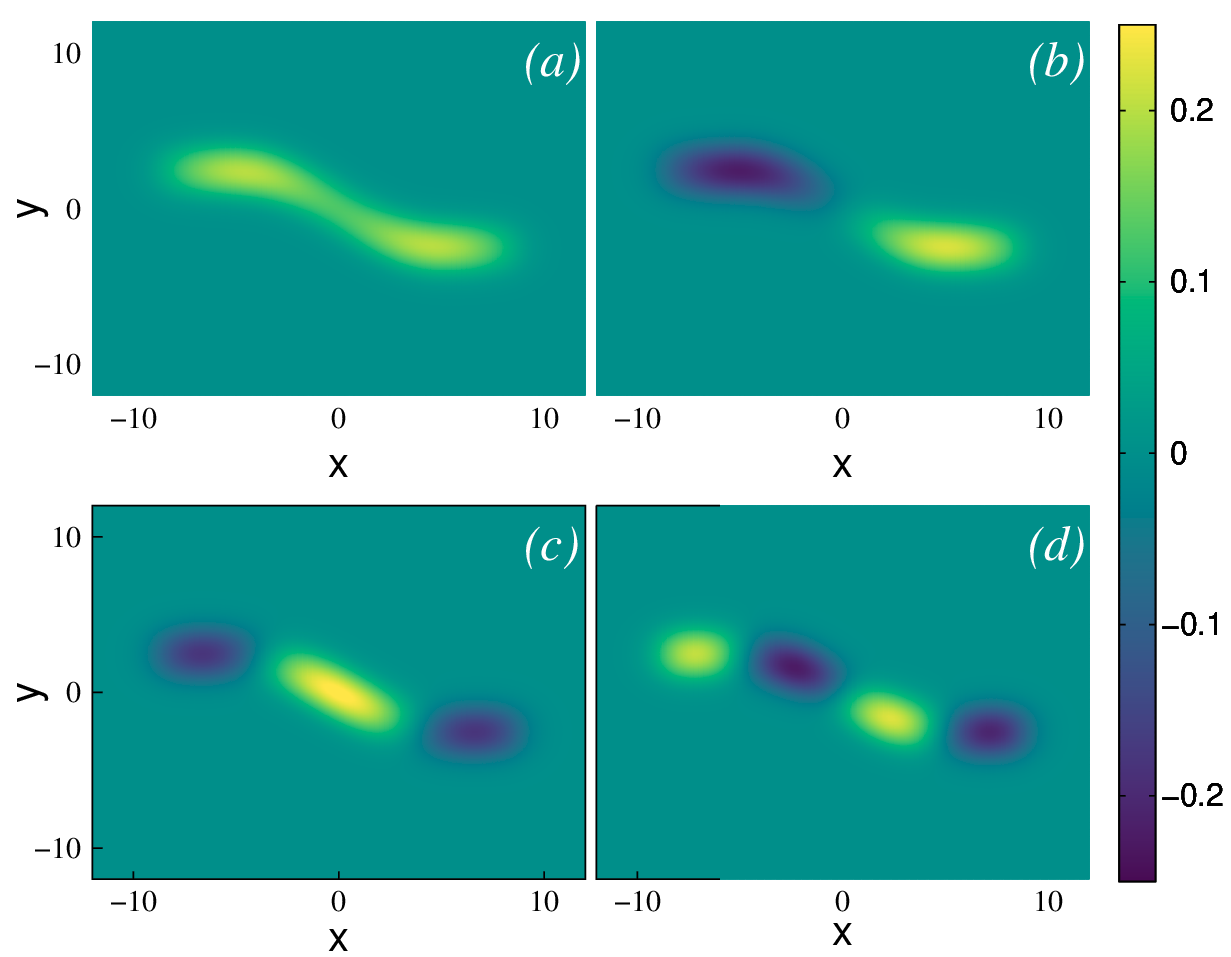}
    \caption{ The first four single-particle eigenstates obtained using exact diagonalization for the bent potential with parameters $B=2.5$, $D=5.0$,$ a=7.0$, and $L=0.5$. Panels ($a$)–($d$) correspond to the ground state and the first, second, and third excited states, respectively. The combination of the ground and first excited states generates left- and right-localized modes. Their energy splitting gives rise to the dominant tunneling frequency observed in the dynamics. The first and second excited states exhibit additional nodal structures on each side, which promote dynamics within individual sides rather than between the two sides. This contributes to the second frequency emerges during the time evolution in the bent potential. All quantities are dimensionless.  }
    \label{figA1}
\end{figure}

To complement the analysis of the two dominant frequencies observed in the dynamics, we examine the spatial structure of the first four eigenstates obtained from ED, see Fig.~\ref{figA1}. These results are presented for $B=2.5$, $D=5.0$, $a=7.0$, and $L=0.5$.
A linear combination of the ground and first excited states [Figs.~\ref{figA1}(a,b)] localize the density on either sides of the bent. The energy splitting between these states sets the primary oscillation frequency in the dynamics, corresponding to the tunneling between left- and right-localized states---that is, full tunneling between the two sides. 
In contrast, the combination of the first and second excited states [Figs.~\ref{figA1}(b,c)] display alternating nodal structures confined to a single side of the bent, representing localized excitations within each side. Thus, their linear combinations are odd within each side, thereby not directly facilitating transport across the bend, but contributing to dynamics within each side. The energy difference between these higher orbitals introduces the secondary frequency in the dynamics (see Table~\ref{tab2}). The resulting interplay between full tunneling across the bent and localized excitations on each side leads to the observed ``breathing" dynamics of $\langle x \rangle$ and $\langle y \rangle$. A full density dynamics video is available in the supplementary material. 

\begin{table}[htbp]
    \centering
    \begin{tabular}{||c|c|c||}
        \hline \hline
        B values & $T= \frac{2 \pi}{(E_2 -E_1)}$ & \makecell{Time period \\ during the dynamics \\ (Numerical)} \\
        \hline
        $B=0.5$ & 159.88  & 160.10  \\ \hline
        $B=1.5$ & 231.15  & 230.92  \\ \hline
        $B=2.5$ & 533.53  & 525.35 \\ \hline
        $B=3.0$ & 992.03  & 1000.66 \\ \hline
        \hline
    \end{tabular}
    \caption{Comparison of the tunneling time periods calculated from the non-interacting energy gap $\left[ T= \frac{2 \pi}{(E_2 -E_1)} \right]$ obtained via exact diagonalization and the time periods extracted from the many-body simulations (MCTDHB) at fixed interaction strength $\Lambda=0.1$, for various half-widths of the bent $B$. All quantities are dimensionless. See Fig.~\ref{fig3} for the dynamics. }
    \label{tab3} 
\end{table}

Table~\ref{tab3} shows a comparison between the tunneling period predicted from the non-interacting energy gap, obtained via exact diagonalization (ED), and the tunneling period computed using the MCTDHB method for weakly interacting bosons. The non-interacting tunneling period is given by $T= \frac{2 \pi}{(E_2 -E_1)}$ where $E_1$ and $E_2$ are the lowest two energies, see Table~\ref{tab1}.

The close agreement between the analytically calculated time period $\left[T= \frac{2 \pi}{(E_2 -E_1)}\right]$ and the numerically obtained oscillation period confirms that the dynamics are dominated by the  two-level tunneling. Nevertheless, the presence of additional ``breathing" motion during the transport (see Fig.~\ref{fig3}) suggests that higher-energy modes also contribute. Consequently, in the many-body calculations, at least $M=4$ orbitals are required to achieve converged results, see the supplementary material for details. These observations imply that the full transport dynamics are more complex and cannot be fully captured by a simple two-mode model.

\bibliographystyle{apsrev4-2-title}  % If available
\bibliography{manuscript}

%apsrev4-2.bst 2019-01-14 (MD) hand-edited version of apsrev4-1.bst
%Control: key (0)
%Control: author (72) initials jnrlst
%Control: editor formatted (1) identically to author
%Control: production of article title (-1) disabled
%Control: page (1) range
%Control: year (1) truncated
%Control: production of eprint (0) enabled
\begin{thebibliography}{62}%
\makeatletter
\providecommand \@ifxundefined [1]{%
 \@ifx{#1\undefined}
}%
\providecommand \@ifnum [1]{%
 \ifnum #1\expandafter \@firstoftwo
 \else \expandafter \@secondoftwo
 \fi
}%
\providecommand \@ifx [1]{%
 \ifx #1\expandafter \@firstoftwo
 \else \expandafter \@secondoftwo
 \fi
}%
\providecommand \natexlab [1]{#1}%
\providecommand \enquote  [1]{``#1''}%
\providecommand \bibnamefont  [1]{#1}%
\providecommand \bibfnamefont [1]{#1}%
\providecommand \citenamefont [1]{#1}%
\providecommand \href@noop [0]{\@secondoftwo}%
\providecommand \href [0]{\begingroup \@sanitize@url \@href}%
\providecommand \@href[1]{\@@startlink{#1}\@@href}%
\providecommand \@@href[1]{\endgroup#1\@@endlink}%
\providecommand \@sanitize@url [0]{\catcode `\\12\catcode `\$12\catcode `\&12\catcode `\#12\catcode `\^12\catcode `\_12\catcode `\%12\relax}%
\providecommand \@@startlink[1]{}%
\providecommand \@@endlink[0]{}%
\providecommand \url  [0]{\begingroup\@sanitize@url \@url }%
\providecommand \@url [1]{\endgroup\@href {#1}{\urlprefix }}%
\providecommand \urlprefix  [0]{URL }%
\providecommand \Eprint [0]{\href }%
\providecommand \doibase [0]{https://doi.org/}%
\providecommand \selectlanguage [0]{\@gobble}%
\providecommand \bibinfo  [0]{\@secondoftwo}%
\providecommand \bibfield  [0]{\@secondoftwo}%
\providecommand \translation [1]{[#1]}%
\providecommand \BibitemOpen [0]{}%
\providecommand \bibitemStop [0]{}%
\providecommand \bibitemNoStop [0]{.\EOS\space}%
\providecommand \EOS [0]{\spacefactor3000\relax}%
\providecommand \BibitemShut  [1]{\csname bibitem#1\endcsname}%
\let\auto@bib@innerbib\@empty
%</preamble>
\bibitem [{\citenamefont {Metcalf}\ \emph {et~al.}(2016)\citenamefont {Metcalf}, \citenamefont {Chern}, \citenamefont {Ventra},\ and\ \citenamefont {Chien}}]{refb1}%
  \BibitemOpen
  \bibfield  {author} {\bibinfo {author} {\bibfnamefont {M.}~\bibnamefont {Metcalf}}, \bibinfo {author} {\bibfnamefont {G.-W.}\ \bibnamefont {Chern}}, \bibinfo {author} {\bibfnamefont {M.~D.}\ \bibnamefont {Ventra}},\ and\ \bibinfo {author} {\bibfnamefont {C.-C.}\ \bibnamefont {Chien}},\ }\bibfield  {title} {\emph {\bibinfo {title} {Matter-wave propagation in optical lattices: geometrical and flat-band effects}},\ }\href@noop {} {\bibfield  {journal} {\bibinfo  {journal} {J. Phys. B At. Mol. Opt. Phys.}\ }\textbf {\bibinfo {volume} {49}},\ \bibinfo {pages} {075301} (\bibinfo {year} {2016})}\BibitemShut {NoStop}%
\bibitem [{\citenamefont {Kaplan}\ \emph {et~al.}(2024)\citenamefont {Kaplan}, \citenamefont {Holder},\ and\ \citenamefont {Yan}}]{refb2}%
  \BibitemOpen
  \bibfield  {author} {\bibinfo {author} {\bibfnamefont {D.}~\bibnamefont {Kaplan}}, \bibinfo {author} {\bibfnamefont {T.}~\bibnamefont {Holder}},\ and\ \bibinfo {author} {\bibfnamefont {B.}~\bibnamefont {Yan}},\ }\bibfield  {title} {\emph {\bibinfo {title} {Unification of nonlinear anomalous hall effect and nonreciprocal magnetoresistance in metals by the quantum geometry}},\ }\href@noop {} {\bibfield  {journal} {\bibinfo  {journal} {Phys. Rev. Lett.}\ }\textbf {\bibinfo {volume} {132}},\ \bibinfo {pages} {026301} (\bibinfo {year} {2024})}\BibitemShut {NoStop}%
\bibitem [{\citenamefont {Mare\ifmmode~\check{s}\else \v{s}\fi{}}\ \emph {et~al.}(2020)\citenamefont {Mare\ifmmode~\check{s}\else \v{s}\fi{}}, \citenamefont {Novotn\'y}, \citenamefont {\ifmmode \check{S}\else \v{S}\fi{}tefa\ifmmode~\check{n}\else \v{n}\fi{}\'ak},\ and\ \citenamefont {Jex}}]{refb3}%
  \BibitemOpen
  \bibfield  {author} {\bibinfo {author} {\bibfnamefont {J.}~\bibnamefont {Mare\ifmmode~\check{s}\else \v{s}\fi{}}}, \bibinfo {author} {\bibfnamefont {J.}~\bibnamefont {Novotn\'y}}, \bibinfo {author} {\bibfnamefont {M.}~\bibnamefont {\ifmmode \check{S}\else \v{S}\fi{}tefa\ifmmode~\check{n}\else \v{n}\fi{}\'ak}},\ and\ \bibinfo {author} {\bibfnamefont {I.}~\bibnamefont {Jex}},\ }\bibfield  {title} {\emph {\bibinfo {title} {Counterintuitive role of geometry in transport by quantum walks}},\ }\href@noop {} {\bibfield  {journal} {\bibinfo  {journal} {Phys. Rev. A}\ }\textbf {\bibinfo {volume} {101}},\ \bibinfo {pages} {032113} (\bibinfo {year} {2020})}\BibitemShut {NoStop}%
\bibitem [{\citenamefont {Kagan}\ and\ \citenamefont {Murray}(2015)}]{refb4}%
  \BibitemOpen
  \bibfield  {author} {\bibinfo {author} {\bibfnamefont {C.~R.}\ \bibnamefont {Kagan}}\ and\ \bibinfo {author} {\bibfnamefont {C.~B.}\ \bibnamefont {Murray}},\ }\bibfield  {title} {\emph {\bibinfo {title} {Charge transport in strongly coupled quantum dot solids}},\ }\href@noop {} {\bibfield  {journal} {\bibinfo  {journal} {Nat. Nanotechnol.}\ }\textbf {\bibinfo {volume} {10}},\ \bibinfo {pages} {1013} (\bibinfo {year} {2015})}\BibitemShut {NoStop}%
\bibitem [{\citenamefont {Minoguchi}\ \emph {et~al.}(2025)\citenamefont {Minoguchi}, \citenamefont {Huber}, \citenamefont {Garbe}, \citenamefont {Gambassi},\ and\ \citenamefont {Rabl}}]{PhysRevLett.134.207102}%
  \BibitemOpen
  \bibfield  {author} {\bibinfo {author} {\bibfnamefont {Y.}~\bibnamefont {Minoguchi}}, \bibinfo {author} {\bibfnamefont {J.}~\bibnamefont {Huber}}, \bibinfo {author} {\bibfnamefont {L.}~\bibnamefont {Garbe}}, \bibinfo {author} {\bibfnamefont {A.}~\bibnamefont {Gambassi}},\ and\ \bibinfo {author} {\bibfnamefont {P.}~\bibnamefont {Rabl}},\ }\bibfield  {title} {\emph {\bibinfo {title} {Unified interface model for dissipative transport of bosons and fermions}},\ }\href@noop {} {\bibfield  {journal} {\bibinfo  {journal} {Phys. Rev. Lett.}\ }\textbf {\bibinfo {volume} {134}},\ \bibinfo {pages} {207102} (\bibinfo {year} {2025})}\BibitemShut {NoStop}%
\bibitem [{\citenamefont {Hartnoll}(2015)}]{Hartnoll2015}%
  \BibitemOpen
  \bibfield  {author} {\bibinfo {author} {\bibfnamefont {S.~A.}\ \bibnamefont {Hartnoll}},\ }\bibfield  {title} {\emph {\bibinfo {title} {Theory of universal incoherent metallic transport}},\ }\href@noop {} {\bibfield  {journal} {\bibinfo  {journal} {Nat. Phys.}\ }\textbf {\bibinfo {volume} {11}},\ \bibinfo {pages} {54} (\bibinfo {year} {2015})}\BibitemShut {NoStop}%
\bibitem [{\citenamefont {Chou}\ \emph {et~al.}(2011)\citenamefont {Chou}, \citenamefont {Mallick},\ and\ \citenamefont {Zia}}]{Chou_2011}%
  \BibitemOpen
  \bibfield  {author} {\bibinfo {author} {\bibfnamefont {T.}~\bibnamefont {Chou}}, \bibinfo {author} {\bibfnamefont {K.}~\bibnamefont {Mallick}},\ and\ \bibinfo {author} {\bibfnamefont {R.~K.~P.}\ \bibnamefont {Zia}},\ }\bibfield  {title} {\emph {\bibinfo {title} {Non-equilibrium statistical mechanics: from a paradigmatic model to biological transport}},\ }\href@noop {} {\bibfield  {journal} {\bibinfo  {journal} {Rep. Prog. Phys.}\ }\textbf {\bibinfo {volume} {74}},\ \bibinfo {pages} {116601} (\bibinfo {year} {2011})}\BibitemShut {NoStop}%
\bibitem [{\citenamefont {Kohler}\ \emph {et~al.}(2005)\citenamefont {Kohler}, \citenamefont {Lehmann},\ and\ \citenamefont {Hänggi}}]{KOHLER2005379}%
  \BibitemOpen
  \bibfield  {author} {\bibinfo {author} {\bibfnamefont {S.}~\bibnamefont {Kohler}}, \bibinfo {author} {\bibfnamefont {J.}~\bibnamefont {Lehmann}},\ and\ \bibinfo {author} {\bibfnamefont {P.}~\bibnamefont {Hänggi}},\ }\bibfield  {title} {\emph {\bibinfo {title} {Driven quantum transport on the nanoscale}},\ }\href@noop {} {\bibfield  {journal} {\bibinfo  {journal} {Phys. Rep.}\ }\textbf {\bibinfo {volume} {406}},\ \bibinfo {pages} {379} (\bibinfo {year} {2005})}\BibitemShut {NoStop}%
\bibitem [{\citenamefont {Tononi}\ \emph {et~al.}(2024)\citenamefont {Tononi}, \citenamefont {Salasnich},\ and\ \citenamefont {Yakimenko}}]{refbb1}%
  \BibitemOpen
  \bibfield  {author} {\bibinfo {author} {\bibfnamefont {A.}~\bibnamefont {Tononi}}, \bibinfo {author} {\bibfnamefont {L.}~\bibnamefont {Salasnich}},\ and\ \bibinfo {author} {\bibfnamefont {A.}~\bibnamefont {Yakimenko}},\ }\bibfield  {title} {\emph {\bibinfo {title} {Quantum vortices in curved geometries}},\ }\href@noop {} {\bibfield  {journal} {\bibinfo  {journal} {AVS Quantum Sci.}\ }\textbf {\bibinfo {volume} {6}},\ \bibinfo {pages} {030502} (\bibinfo {year} {2024})}\BibitemShut {NoStop}%
\bibitem [{\citenamefont {M.~Pitelli}\ \emph {et~al.}(2024)\citenamefont {M.~Pitelli}, \citenamefont {Mosna},\ and\ \citenamefont {Felix~Souto}}]{refbb3}%
  \BibitemOpen
  \bibfield  {author} {\bibinfo {author} {\bibfnamefont {J.~P.}\ \bibnamefont {M.~Pitelli}}, \bibinfo {author} {\bibfnamefont {R.~A.}\ \bibnamefont {Mosna}},\ and\ \bibinfo {author} {\bibfnamefont {F.}~\bibnamefont {Felix~Souto}},\ }\bibfield  {title} {\emph {\bibinfo {title} {Quantum mechanics on sharply bent wires via two-interval {Sturm-Liouville} theory}},\ }\href@noop {} {\bibfield  {journal} {\bibinfo  {journal} {J. Math. Phys.}\ }\textbf {\bibinfo {volume} {65}},\ \bibinfo {pages} {062104} (\bibinfo {year} {2024})}\BibitemShut {NoStop}%
\bibitem [{\citenamefont {Marchi}\ \emph {et~al.}(2005)\citenamefont {Marchi}, \citenamefont {Reggiani}, \citenamefont {Rudan},\ and\ \citenamefont {Bertoni}}]{refbe1}%
  \BibitemOpen
  \bibfield  {author} {\bibinfo {author} {\bibfnamefont {A.}~\bibnamefont {Marchi}}, \bibinfo {author} {\bibfnamefont {S.}~\bibnamefont {Reggiani}}, \bibinfo {author} {\bibfnamefont {M.}~\bibnamefont {Rudan}},\ and\ \bibinfo {author} {\bibfnamefont {A.}~\bibnamefont {Bertoni}},\ }\bibfield  {title} {\emph {\bibinfo {title} {Coherent electron transport in bent cylindrical surfaces}},\ }\href@noop {} {\bibfield  {journal} {\bibinfo  {journal} {Phys. Rev. B}\ }\textbf {\bibinfo {volume} {72}},\ \bibinfo {pages} {035403} (\bibinfo {year} {2005})}\BibitemShut {NoStop}%
\bibitem [{\citenamefont {Du}\ \emph {et~al.}(2016)\citenamefont {Du}, \citenamefont {Wang}, \citenamefont {Liang}, \citenamefont {Kang}, \citenamefont {Liu},\ and\ \citenamefont {Zong}}]{refbe2}%
  \BibitemOpen
  \bibfield  {author} {\bibinfo {author} {\bibfnamefont {L.}~\bibnamefont {Du}}, \bibinfo {author} {\bibfnamefont {Y.-L.}\ \bibnamefont {Wang}}, \bibinfo {author} {\bibfnamefont {G.-H.}\ \bibnamefont {Liang}}, \bibinfo {author} {\bibfnamefont {G.-Z.}\ \bibnamefont {Kang}}, \bibinfo {author} {\bibfnamefont {X.-J.}\ \bibnamefont {Liu}},\ and\ \bibinfo {author} {\bibfnamefont {H.-S.}\ \bibnamefont {Zong}},\ }\bibfield  {title} {\emph {\bibinfo {title} {Curvature-induced bound states and coherent electron transport on the surface of a truncated cone}},\ }\href@noop {} {\bibfield  {journal} {\bibinfo  {journal} {Physica E}\ }\textbf {\bibinfo {volume} {76}},\ \bibinfo {pages} {28} (\bibinfo {year} {2016})}\BibitemShut {NoStop}%
\bibitem [{\citenamefont {Campo}\ \emph {et~al.}(2014)\citenamefont {Campo}, \citenamefont {Boshier},\ and\ \citenamefont {Saxena}}]{refbd1}%
  \BibitemOpen
  \bibfield  {author} {\bibinfo {author} {\bibfnamefont {A.~d.}\ \bibnamefont {Campo}}, \bibinfo {author} {\bibfnamefont {M.~G.}\ \bibnamefont {Boshier}},\ and\ \bibinfo {author} {\bibfnamefont {A.}~\bibnamefont {Saxena}},\ }\bibfield  {title} {\emph {\bibinfo {title} {Bent waveguides for matter-waves: supersymmetric potentials and reflectionless geometries}},\ }\href@noop {} {\bibfield  {journal} {\bibinfo  {journal} {Sci. Rep.}\ }\textbf {\bibinfo {volume} {4}},\ \bibinfo {pages} {5274} (\bibinfo {year} {2014})}\BibitemShut {NoStop}%
\bibitem [{\citenamefont {Ryu}\ and\ \citenamefont {Boshier}(2015{\natexlab{a}})}]{refbd2}%
  \BibitemOpen
  \bibfield  {author} {\bibinfo {author} {\bibfnamefont {C.}~\bibnamefont {Ryu}}\ and\ \bibinfo {author} {\bibfnamefont {M.~G.}\ \bibnamefont {Boshier}},\ }\bibfield  {title} {\emph {\bibinfo {title} {Integrated coherent matter wave circuits}},\ }\href@noop {} {\bibfield  {journal} {\bibinfo  {journal} {New J. Phys.}\ }\textbf {\bibinfo {volume} {17}},\ \bibinfo {pages} {092002} (\bibinfo {year} {2015}{\natexlab{a}})}\BibitemShut {NoStop}%
\bibitem [{\citenamefont {Kleinherbers}\ \emph {et~al.}(2023)\citenamefont {Kleinherbers}, \citenamefont {Stegmann},\ and\ \citenamefont {Szpak}}]{refbd3}%
  \BibitemOpen
  \bibfield  {author} {\bibinfo {author} {\bibfnamefont {E.}~\bibnamefont {Kleinherbers}}, \bibinfo {author} {\bibfnamefont {T.}~\bibnamefont {Stegmann}},\ and\ \bibinfo {author} {\bibfnamefont {N.}~\bibnamefont {Szpak}},\ }\bibfield  {title} {\emph {\bibinfo {title} {Electronic transport in bent carbon nanotubes}},\ }\href@noop {} {\bibfield  {journal} {\bibinfo  {journal} {Phys. Rev. B}\ }\textbf {\bibinfo {volume} {107}},\ \bibinfo {pages} {195424} (\bibinfo {year} {2023})}\BibitemShut {NoStop}%
\bibitem [{\citenamefont {Amico}\ \emph {et~al.}(2022)\citenamefont {Amico}, \citenamefont {Anderson}, \citenamefont {Boshier}, \citenamefont {Brantut}, \citenamefont {Kwek}, \citenamefont {Minguzzi},\ and\ \citenamefont {von Klitzing}}]{refbd4}%
  \BibitemOpen
  \bibfield  {author} {\bibinfo {author} {\bibfnamefont {L.}~\bibnamefont {Amico}}, \bibinfo {author} {\bibfnamefont {D.}~\bibnamefont {Anderson}}, \bibinfo {author} {\bibfnamefont {M.}~\bibnamefont {Boshier}}, \bibinfo {author} {\bibfnamefont {J.-P.}\ \bibnamefont {Brantut}}, \bibinfo {author} {\bibfnamefont {L.-C.}\ \bibnamefont {Kwek}}, \bibinfo {author} {\bibfnamefont {A.}~\bibnamefont {Minguzzi}},\ and\ \bibinfo {author} {\bibfnamefont {W.}~\bibnamefont {von Klitzing}},\ }\bibfield  {title} {\emph {\bibinfo {title} {Colloquium: Atomtronic circuits: From many-body physics to quantum technologies}},\ }\href@noop {} {\bibfield  {journal} {\bibinfo  {journal} {Rev. Mod. Phys.}\ }\textbf {\bibinfo {volume} {94}},\ \bibinfo {pages} {041001} (\bibinfo {year} {2022})}\BibitemShut {NoStop}%
\bibitem [{\citenamefont {Rhyno}\ \emph {et~al.}(2021)\citenamefont {Rhyno}, \citenamefont {Lundblad}, \citenamefont {Aveline}, \citenamefont {Lannert},\ and\ \citenamefont {Vishveshwara}}]{ref2b1}%
  \BibitemOpen
  \bibfield  {author} {\bibinfo {author} {\bibfnamefont {B.}~\bibnamefont {Rhyno}}, \bibinfo {author} {\bibfnamefont {N.}~\bibnamefont {Lundblad}}, \bibinfo {author} {\bibfnamefont {D.~C.}\ \bibnamefont {Aveline}}, \bibinfo {author} {\bibfnamefont {C.}~\bibnamefont {Lannert}},\ and\ \bibinfo {author} {\bibfnamefont {S.}~\bibnamefont {Vishveshwara}},\ }\bibfield  {title} {\emph {\bibinfo {title} {Thermodynamics in expanding shell-shaped {Bose-Einstein} condensates}},\ }\href@noop {} {\bibfield  {journal} {\bibinfo  {journal} {Phys. Rev. A}\ }\textbf {\bibinfo {volume} {104}},\ \bibinfo {pages} {063310} (\bibinfo {year} {2021})}\BibitemShut {NoStop}%
\bibitem [{\citenamefont {Diniz}\ \emph {et~al.}(2020)\citenamefont {Diniz}, \citenamefont {Oliveira}, \citenamefont {Lima},\ and\ \citenamefont {Henn}}]{ref2b1a}%
  \BibitemOpen
  \bibfield  {author} {\bibinfo {author} {\bibfnamefont {P.~C.}\ \bibnamefont {Diniz}}, \bibinfo {author} {\bibfnamefont {E.~A.~B.}\ \bibnamefont {Oliveira}}, \bibinfo {author} {\bibfnamefont {A.~R.~P.}\ \bibnamefont {Lima}},\ and\ \bibinfo {author} {\bibfnamefont {E.~A.~L.}\ \bibnamefont {Henn}},\ }\bibfield  {title} {\emph {\bibinfo {title} {Ground state and collective excitations of a dipolar {Bose-Einstein} condensate in a bubble trap}},\ }\href@noop {} {\bibfield  {journal} {\bibinfo  {journal} {Sci. Rep.}\ }\textbf {\bibinfo {volume} {10}},\ \bibinfo {pages} {4831} (\bibinfo {year} {2020})}\BibitemShut {NoStop}%
\bibitem [{\citenamefont {Padavi\ifmmode~\acute{c}\else \'{c}\fi{}}\ \emph {et~al.}(2020)\citenamefont {Padavi\ifmmode~\acute{c}\else \'{c}\fi{}}, \citenamefont {Sun}, \citenamefont {Lannert},\ and\ \citenamefont {Vishveshwara}}]{ref2b2}%
  \BibitemOpen
  \bibfield  {author} {\bibinfo {author} {\bibfnamefont {K.}~\bibnamefont {Padavi\ifmmode~\acute{c}\else \'{c}\fi{}}}, \bibinfo {author} {\bibfnamefont {K.}~\bibnamefont {Sun}}, \bibinfo {author} {\bibfnamefont {C.}~\bibnamefont {Lannert}},\ and\ \bibinfo {author} {\bibfnamefont {S.}~\bibnamefont {Vishveshwara}},\ }\bibfield  {title} {\emph {\bibinfo {title} {Vortex-antivortex physics in shell-shaped {Bose-Einstein} condensates}},\ }\href@noop {} {\bibfield  {journal} {\bibinfo  {journal} {Phys. Rev. A}\ }\textbf {\bibinfo {volume} {102}},\ \bibinfo {pages} {043305} (\bibinfo {year} {2020})}\BibitemShut {NoStop}%
\bibitem [{\citenamefont {Guenther}\ \emph {et~al.}(2017)\citenamefont {Guenther}, \citenamefont {Massignan},\ and\ \citenamefont {Fetter}}]{ref2b3}%
  \BibitemOpen
  \bibfield  {author} {\bibinfo {author} {\bibfnamefont {N.-E.}\ \bibnamefont {Guenther}}, \bibinfo {author} {\bibfnamefont {P.}~\bibnamefont {Massignan}},\ and\ \bibinfo {author} {\bibfnamefont {A.~L.}\ \bibnamefont {Fetter}},\ }\bibfield  {title} {\emph {\bibinfo {title} {Quantized superfluid vortex dynamics on cylindrical surfaces and planar annuli}},\ }\href@noop {} {\bibfield  {journal} {\bibinfo  {journal} {Phys. Rev. A}\ }\textbf {\bibinfo {volume} {96}},\ \bibinfo {pages} {063608} (\bibinfo {year} {2017})}\BibitemShut {NoStop}%
\bibitem [{\citenamefont {Bereta}\ \emph {et~al.}(2021)\citenamefont {Bereta}, \citenamefont {Caracanhas},\ and\ \citenamefont {Fetter}}]{ref2b4}%
  \BibitemOpen
  \bibfield  {author} {\bibinfo {author} {\bibfnamefont {S.~J.}\ \bibnamefont {Bereta}}, \bibinfo {author} {\bibfnamefont {M.~A.}\ \bibnamefont {Caracanhas}},\ and\ \bibinfo {author} {\bibfnamefont {A.~L.}\ \bibnamefont {Fetter}},\ }\bibfield  {title} {\emph {\bibinfo {title} {Superfluid vortex dynamics on a spherical film}},\ }\href@noop {} {\bibfield  {journal} {\bibinfo  {journal} {Phys. Rev. A}\ }\textbf {\bibinfo {volume} {103}},\ \bibinfo {pages} {053306} (\bibinfo {year} {2021})}\BibitemShut {NoStop}%
\bibitem [{\citenamefont {Ryu}\ and\ \citenamefont {Boshier}(2015{\natexlab{b}})}]{bent_exp1}%
  \BibitemOpen
  \bibfield  {author} {\bibinfo {author} {\bibfnamefont {C.}~\bibnamefont {Ryu}}\ and\ \bibinfo {author} {\bibfnamefont {M.~G.}\ \bibnamefont {Boshier}},\ }\bibfield  {title} {\emph {\bibinfo {title} {Integrated coherent matter wave circuits}},\ }\href@noop {} {\bibfield  {journal} {\bibinfo  {journal} {New J. Phys.}\ }\textbf {\bibinfo {volume} {17}},\ \bibinfo {pages} {092002} (\bibinfo {year} {2015}{\natexlab{b}})}\BibitemShut {NoStop}%
\bibitem [{\citenamefont {Henderson}\ \emph {et~al.}(2009)\citenamefont {Henderson}, \citenamefont {Ryu}, \citenamefont {MacCormick},\ and\ \citenamefont {Boshier}}]{bent_exp2}%
  \BibitemOpen
  \bibfield  {author} {\bibinfo {author} {\bibfnamefont {K.}~\bibnamefont {Henderson}}, \bibinfo {author} {\bibfnamefont {C.}~\bibnamefont {Ryu}}, \bibinfo {author} {\bibfnamefont {C.}~\bibnamefont {MacCormick}},\ and\ \bibinfo {author} {\bibfnamefont {M.~G.}\ \bibnamefont {Boshier}},\ }\bibfield  {title} {\emph {\bibinfo {title} {Experimental demonstration of painting arbitrary and dynamic potentials for {Bose–Einstein} condensates}},\ }\href@noop {} {\bibfield  {journal} {\bibinfo  {journal} {New J. Phys.}\ }\textbf {\bibinfo {volume} {11}},\ \bibinfo {pages} {043030} (\bibinfo {year} {2009})}\BibitemShut {NoStop}%
\bibitem [{\citenamefont {Amico}\ \emph {et~al.}(2021)\citenamefont {Amico}, \citenamefont {Boshier}, \citenamefont {Birkl}, \citenamefont {Minguzzi},\ and\ \citenamefont {\textit{et al.}}}]{bent_exp3}%
  \BibitemOpen
  \bibfield  {author} {\bibinfo {author} {\bibfnamefont {L.}~\bibnamefont {Amico}}, \bibinfo {author} {\bibfnamefont {M.}~\bibnamefont {Boshier}}, \bibinfo {author} {\bibfnamefont {G.}~\bibnamefont {Birkl}}, \bibinfo {author} {\bibfnamefont {A.}~\bibnamefont {Minguzzi}},\ and\ \bibinfo {author} {\bibnamefont {\textit{et al.}}},\ }\bibfield  {title} {\emph {\bibinfo {title} {Roadmap on atomtronics: State of the art and perspective}},\ }\href@noop {} {\bibfield  {journal} {\bibinfo  {journal} {AVS Quantum Sci.}\ }\textbf {\bibinfo {volume} {3}},\ \bibinfo {pages} {039201} (\bibinfo {year} {2021})}\BibitemShut {NoStop}%
\bibitem [{\citenamefont {Naldesi}\ \emph {et~al.}(2022)\citenamefont {Naldesi}, \citenamefont {Polo}, \citenamefont {Dunjko}, \citenamefont {Perrin}, \citenamefont {Olshanii}, \citenamefont {Amico},\ and\ \citenamefont {Minguzzi}}]{bent_exp4}%
  \BibitemOpen
  \bibfield  {author} {\bibinfo {author} {\bibfnamefont {P.}~\bibnamefont {Naldesi}}, \bibinfo {author} {\bibfnamefont {J.}~\bibnamefont {Polo}}, \bibinfo {author} {\bibfnamefont {V.}~\bibnamefont {Dunjko}}, \bibinfo {author} {\bibfnamefont {H.}~\bibnamefont {Perrin}}, \bibinfo {author} {\bibfnamefont {M.}~\bibnamefont {Olshanii}}, \bibinfo {author} {\bibfnamefont {L.}~\bibnamefont {Amico}},\ and\ \bibinfo {author} {\bibfnamefont {A.}~\bibnamefont {Minguzzi}},\ }\bibfield  {title} {\emph {\bibinfo {title} {{Enhancing sensitivity to rotations with quantum solitonic currents}}},\ }\href@noop {} {\bibfield  {journal} {\bibinfo  {journal} {SciPost Phys.}\ }\textbf {\bibinfo {volume} {12}},\ \bibinfo {pages} {138} (\bibinfo {year} {2022})}\BibitemShut {NoStop}%
\bibitem [{\citenamefont {Duclos}\ and\ \citenamefont {Exner}(1995)}]{S0129055X95000062}%
  \BibitemOpen
  \bibfield  {author} {\bibinfo {author} {\bibfnamefont {P.}~\bibnamefont {Duclos}}\ and\ \bibinfo {author} {\bibfnamefont {P.}~\bibnamefont {Exner}},\ }\bibfield  {title} {\emph {\bibinfo {title} {Curvature-induced bound states in quantum waveguides in two and three dimensions}},\ }\href@noop {} {\bibfield  {journal} {\bibinfo  {journal} {Rev. Math. Phys.}\ }\textbf {\bibinfo {volume} {07}},\ \bibinfo {pages} {73} (\bibinfo {year} {1995})}\BibitemShut {NoStop}%
\bibitem [{\citenamefont {Bittner}\ \emph {et~al.}(2013)\citenamefont {Bittner}, \citenamefont {Dietz}, \citenamefont {Miski-Oglu}, \citenamefont {Richter}, \citenamefont {Ripp}, \citenamefont {Sadurn\'{\i}},\ and\ \citenamefont {Schleich}}]{PhysRevE.87.042912}%
  \BibitemOpen
  \bibfield  {author} {\bibinfo {author} {\bibfnamefont {S.}~\bibnamefont {Bittner}}, \bibinfo {author} {\bibfnamefont {B.}~\bibnamefont {Dietz}}, \bibinfo {author} {\bibfnamefont {M.}~\bibnamefont {Miski-Oglu}}, \bibinfo {author} {\bibfnamefont {A.}~\bibnamefont {Richter}}, \bibinfo {author} {\bibfnamefont {C.}~\bibnamefont {Ripp}}, \bibinfo {author} {\bibfnamefont {E.}~\bibnamefont {Sadurn\'{\i}}},\ and\ \bibinfo {author} {\bibfnamefont {W.~P.}\ \bibnamefont {Schleich}},\ }\bibfield  {title} {\emph {\bibinfo {title} {Bound states in sharply bent waveguides: Analytical and experimental approach}},\ }\href@noop {} {\bibfield  {journal} {\bibinfo  {journal} {Phys. Rev. E}\ }\textbf {\bibinfo {volume} {87}},\ \bibinfo {pages} {042912} (\bibinfo {year} {2013})}\BibitemShut {NoStop}%
\bibitem [{\citenamefont {Haldar}\ and\ \citenamefont {Alon}(2019)}]{sudip_asymmetric_dw}%
  \BibitemOpen
  \bibfield  {author} {\bibinfo {author} {\bibfnamefont {S.~K.}\ \bibnamefont {Haldar}}\ and\ \bibinfo {author} {\bibfnamefont {O.~E.}\ \bibnamefont {Alon}},\ }\bibfield  {title} {\emph {\bibinfo {title} {Many-body quantum dynamics of an asymmetric bosonic {Josephson} junction}},\ }\href@noop {} {\bibfield  {journal} {\bibinfo  {journal} {New J. Phys.}\ }\textbf {\bibinfo {volume} {21}},\ \bibinfo {pages} {103037} (\bibinfo {year} {2019})}\BibitemShut {NoStop}%
\bibitem [{\citenamefont {Bhowmik}\ \emph {et~al.}(2020)\citenamefont {Bhowmik}, \citenamefont {Haldar},\ and\ \citenamefont {Alon}}]{anal_2020_dw}%
  \BibitemOpen
  \bibfield  {author} {\bibinfo {author} {\bibfnamefont {A.}~\bibnamefont {Bhowmik}}, \bibinfo {author} {\bibfnamefont {S.~K.}\ \bibnamefont {Haldar}},\ and\ \bibinfo {author} {\bibfnamefont {O.~E.}\ \bibnamefont {Alon}},\ }\bibfield  {title} {\emph {\bibinfo {title} {Impact of the transverse direction on the many-body tunneling dynamics in a two-dimensional bosonic {Josephson} junction}},\ }\href@noop {} {\bibfield  {journal} {\bibinfo  {journal} {Sci. Rep.}\ }\textbf {\bibinfo {volume} {10}},\ \bibinfo {pages} {21476} (\bibinfo {year} {2020})}\BibitemShut {NoStop}%
\bibitem [{\citenamefont {Streltsov}\ \emph {et~al.}(2007)\citenamefont {Streltsov}, \citenamefont {Alon},\ and\ \citenamefont {Cederbaum}}]{MCTDHB2}%
  \BibitemOpen
  \bibfield  {author} {\bibinfo {author} {\bibfnamefont {A.~I.}\ \bibnamefont {Streltsov}}, \bibinfo {author} {\bibfnamefont {O.~E.}\ \bibnamefont {Alon}},\ and\ \bibinfo {author} {\bibfnamefont {L.~S.}\ \bibnamefont {Cederbaum}},\ }\bibfield  {title} {\emph {\bibinfo {title} {Role of excited states in the splitting of a trapped interacting {Bose-Einstein} condensate by a time-dependent barrier}},\ }\href@noop {} {\bibfield  {journal} {\bibinfo  {journal} {Phys. Rev. Lett.}\ }\textbf {\bibinfo {volume} {99}},\ \bibinfo {pages} {030402} (\bibinfo {year} {2007})}\BibitemShut {NoStop}%
\bibitem [{\citenamefont {Roy}\ and\ \citenamefont {Alon}(2025)}]{rhombik_acc}%
  \BibitemOpen
  \bibfield  {author} {\bibinfo {author} {\bibfnamefont {R.}~\bibnamefont {Roy}}\ and\ \bibinfo {author} {\bibfnamefont {O.~E.}\ \bibnamefont {Alon}},\ }\bibfield  {title} {\emph {\bibinfo {title} {Assessing small accelerations using a bosonic {Josephson} junction}},\ }\href@noop {} {\bibfield  {journal} {\bibinfo  {journal} {Phys. Rev. A}\ }\textbf {\bibinfo {volume} {111}},\ \bibinfo {pages} {043307} (\bibinfo {year} {2025})}\BibitemShut {NoStop}%
\bibitem [{\citenamefont {Alon}\ \emph {et~al.}(2008)\citenamefont {Alon}, \citenamefont {Streltsov},\ and\ \citenamefont {Cederbaum}}]{MCTDHB1}%
  \BibitemOpen
  \bibfield  {author} {\bibinfo {author} {\bibfnamefont {O.~E.}\ \bibnamefont {Alon}}, \bibinfo {author} {\bibfnamefont {A.~I.}\ \bibnamefont {Streltsov}},\ and\ \bibinfo {author} {\bibfnamefont {L.~S.}\ \bibnamefont {Cederbaum}},\ }\bibfield  {title} {\emph {\bibinfo {title} {Multiconfigurational time-dependent {Hartree} method for bosons: Many-body dynamics of bosonic systems}},\ }\href@noop {} {\bibfield  {journal} {\bibinfo  {journal} {Phys. Rev. A}\ }\textbf {\bibinfo {volume} {77}},\ \bibinfo {pages} {033613} (\bibinfo {year} {2008})}\BibitemShut {NoStop}%
\bibitem [{\citenamefont {Alon}\ \emph {et~al.}(2007)\citenamefont {Alon}, \citenamefont {Streltsov},\ and\ \citenamefont {Cederbaum}}]{mctdhb_exact2}%
  \BibitemOpen
  \bibfield  {author} {\bibinfo {author} {\bibfnamefont {O.~E.}\ \bibnamefont {Alon}}, \bibinfo {author} {\bibfnamefont {A.~I.}\ \bibnamefont {Streltsov}},\ and\ \bibinfo {author} {\bibfnamefont {L.~S.}\ \bibnamefont {Cederbaum}},\ }\bibfield  {title} {\emph {\bibinfo {title} {Unified view on multiconfigurational time propagation for systems consisting of identical particles}},\ }\href@noop {} {\bibfield  {journal} {\bibinfo  {journal} {J. Chem. Phys.}\ }\textbf {\bibinfo {volume} {127}},\ \bibinfo {pages} {154103} (\bibinfo {year} {2007})}\BibitemShut {NoStop}%
\bibitem [{\citenamefont {Kramer}\ and\ \citenamefont {Saraceno}(1981)}]{variational5}%
  \BibitemOpen
  \bibfield  {author} {\bibinfo {author} {\bibfnamefont {P.}~\bibnamefont {Kramer}}\ and\ \bibinfo {author} {\bibfnamefont {M.}~\bibnamefont {Saraceno}},\ }\href@noop {} {\emph {\bibinfo {title} {Geometry of the Time-Dependent Variational Principle in Quantum Mechanics}}}\ (\bibinfo  {publisher} {Springer Berlin Heidelberg},\ \bibinfo {address} {Berlin, Heidelberg},\ \bibinfo {year} {1981})\BibitemShut {NoStop}%
\bibitem [{\citenamefont {Lode}\ \emph {et~al.}(2024)\citenamefont {Lode}, \citenamefont {Tsatsos}, \citenamefont {Fasshauer}, \citenamefont {Weiner}, \citenamefont {Lin}, \citenamefont {Papariello}, \citenamefont {Molignini}, \citenamefont {L{\'{e}}v{\^{e}}que}, \citenamefont {B\"uttner}, \citenamefont {Xiang}, \citenamefont {Dutta}, \citenamefont {Roy}, \citenamefont {Bilinskaya},\ and\ \citenamefont {Eder}}]{mctdhb_software1}%
  \BibitemOpen
  \bibfield  {author} {\bibinfo {author} {\bibfnamefont {A.~U.~J.}\ \bibnamefont {Lode}}, \bibinfo {author} {\bibfnamefont {M.~C.}\ \bibnamefont {Tsatsos}}, \bibinfo {author} {\bibfnamefont {E.}~\bibnamefont {Fasshauer}}, \bibinfo {author} {\bibfnamefont {S.~E.}\ \bibnamefont {Weiner}}, \bibinfo {author} {\bibfnamefont {R.}~\bibnamefont {Lin}}, \bibinfo {author} {\bibfnamefont {L.}~\bibnamefont {Papariello}}, \bibinfo {author} {\bibfnamefont {P.}~\bibnamefont {Molignini}}, \bibinfo {author} {\bibfnamefont {C.}~\bibnamefont {L{\'{e}}v{\^{e}}que}}, \bibinfo {author} {\bibfnamefont {M.}~\bibnamefont {B\"uttner}}, \bibinfo {author} {\bibfnamefont {J.}~\bibnamefont {Xiang}}, \bibinfo {author} {\bibfnamefont {S.}~\bibnamefont {Dutta}}, \bibinfo {author} {\bibfnamefont {R.}~\bibnamefont {Roy}}, \bibinfo {author} {\bibfnamefont {Y.}~\bibnamefont {Bilinskaya}},\ and\ \bibinfo {author} {\bibfnamefont {M.}~\bibnamefont {Eder}},\ }\href@noop {} {\bibinfo {title} {{MCTDH-X}: The multiconfigurational time-dependent hartree
  method for indistinguishable particles software}} (\bibinfo {year} {2024})\BibitemShut {NoStop}%
\bibitem [{\citenamefont {Lin}\ \emph {et~al.}(2020{\natexlab{a}})\citenamefont {Lin}, \citenamefont {Molignini}, \citenamefont {Papariello}, \citenamefont {Tsatsos}, \citenamefont {L{\'{e}}v{\^{e}}que}, \citenamefont {Weiner}, \citenamefont {Fasshauer}, \citenamefont {Chitra},\ and\ \citenamefont {Lode}}]{mctdhb_software2}%
  \BibitemOpen
  \bibfield  {author} {\bibinfo {author} {\bibfnamefont {R.}~\bibnamefont {Lin}}, \bibinfo {author} {\bibfnamefont {P.}~\bibnamefont {Molignini}}, \bibinfo {author} {\bibfnamefont {L.}~\bibnamefont {Papariello}}, \bibinfo {author} {\bibfnamefont {M.~C.}\ \bibnamefont {Tsatsos}}, \bibinfo {author} {\bibfnamefont {C.}~\bibnamefont {L{\'{e}}v{\^{e}}que}}, \bibinfo {author} {\bibfnamefont {S.~E.}\ \bibnamefont {Weiner}}, \bibinfo {author} {\bibfnamefont {E.}~\bibnamefont {Fasshauer}}, \bibinfo {author} {\bibfnamefont {R.}~\bibnamefont {Chitra}},\ and\ \bibinfo {author} {\bibfnamefont {A.~U.~J.}\ \bibnamefont {Lode}},\ }\bibfield  {title} {\emph {\bibinfo {title} {{MCTDH}-x: The multiconfigurational time-dependent {Hartree} method for indistinguishable particles software}},\ }\href@noop {} {\bibfield  {journal} {\bibinfo  {journal} {Quant. Sci. Technol.}\ }\textbf {\bibinfo {volume} {5}},\ \bibinfo {pages} {024004} (\bibinfo {year} {2020}{\natexlab{a}})}\BibitemShut {NoStop}%
\bibitem [{\citenamefont {Lode}\ \emph {et~al.}(2020)\citenamefont {Lode}, \citenamefont {L\'ev\^eque}, \citenamefont {Madsen}, \citenamefont {Streltsov},\ and\ \citenamefont {Alon}}]{mctdhb_review}%
  \BibitemOpen
  \bibfield  {author} {\bibinfo {author} {\bibfnamefont {A.~U.~J.}\ \bibnamefont {Lode}}, \bibinfo {author} {\bibfnamefont {C.}~\bibnamefont {L\'ev\^eque}}, \bibinfo {author} {\bibfnamefont {L.~B.}\ \bibnamefont {Madsen}}, \bibinfo {author} {\bibfnamefont {A.~I.}\ \bibnamefont {Streltsov}},\ and\ \bibinfo {author} {\bibfnamefont {O.~E.}\ \bibnamefont {Alon}},\ }\bibfield  {title} {\emph {\bibinfo {title} {Colloquium: Multiconfigurational time-dependent {Hartree} approaches for indistinguishable particles}},\ }\href@noop {} {\bibfield  {journal} {\bibinfo  {journal} {Rev. Mod. Phys.}\ }\textbf {\bibinfo {volume} {92}},\ \bibinfo {pages} {011001} (\bibinfo {year} {2020})}\BibitemShut {NoStop}%
\bibitem [{\citenamefont {Roy}\ \emph {et~al.}(2025)\citenamefont {Roy}, \citenamefont {Dutta},\ and\ \citenamefont {Alon}}]{rhombik_scirep}%
  \BibitemOpen
  \bibfield  {author} {\bibinfo {author} {\bibfnamefont {R.}~\bibnamefont {Roy}}, \bibinfo {author} {\bibfnamefont {S.}~\bibnamefont {Dutta}},\ and\ \bibinfo {author} {\bibfnamefont {O.~E.}\ \bibnamefont {Alon}},\ }\bibfield  {title} {\emph {\bibinfo {title} {Rotation quenches in trapped bosonic systems}},\ }\href@noop {} {\bibfield  {journal} {\bibinfo  {journal} {Sci. Rep.}\ }\textbf {\bibinfo {volume} {15}},\ \bibinfo {pages} {27193} (\bibinfo {year} {2025})}\BibitemShut {NoStop}%
\bibitem [{\citenamefont {Lin}\ \emph {et~al.}(2020{\natexlab{b}})\citenamefont {Lin}, \citenamefont {Molignini}, \citenamefont {Lode},\ and\ \citenamefont {Chitra}}]{paolo_cavity2}%
  \BibitemOpen
  \bibfield  {author} {\bibinfo {author} {\bibfnamefont {R.}~\bibnamefont {Lin}}, \bibinfo {author} {\bibfnamefont {P.}~\bibnamefont {Molignini}}, \bibinfo {author} {\bibfnamefont {A.~U.~J.}\ \bibnamefont {Lode}},\ and\ \bibinfo {author} {\bibfnamefont {R.}~\bibnamefont {Chitra}},\ }\bibfield  {title} {\emph {\bibinfo {title} {Pathway to chaos through hierarchical superfluidity in blue-detuned cavity-{BEC} systems}},\ }\href@noop {} {\bibfield  {journal} {\bibinfo  {journal} {Phys. Rev. A}\ }\textbf {\bibinfo {volume} {101}},\ \bibinfo {pages} {061602} (\bibinfo {year} {2020}{\natexlab{b}})}\BibitemShut {NoStop}%
\bibitem [{\citenamefont {Molignini}\ \emph {et~al.}(2022)\citenamefont {Molignini}, \citenamefont {L\'ev\^eque}, \citenamefont {Ke\ss{}ler}, \citenamefont {Jaksch}, \citenamefont {Chitra},\ and\ \citenamefont {Lode}}]{paolo_cz}%
  \BibitemOpen
  \bibfield  {author} {\bibinfo {author} {\bibfnamefont {P.}~\bibnamefont {Molignini}}, \bibinfo {author} {\bibfnamefont {C.}~\bibnamefont {L\'ev\^eque}}, \bibinfo {author} {\bibfnamefont {H.}~\bibnamefont {Ke\ss{}ler}}, \bibinfo {author} {\bibfnamefont {D.}~\bibnamefont {Jaksch}}, \bibinfo {author} {\bibfnamefont {R.}~\bibnamefont {Chitra}},\ and\ \bibinfo {author} {\bibfnamefont {A.~U.~J.}\ \bibnamefont {Lode}},\ }\bibfield  {title} {\emph {\bibinfo {title} {Crystallization via cavity-assisted infinite-range interactions}},\ }\href@noop {} {\bibfield  {journal} {\bibinfo  {journal} {Phys. Rev. A}\ }\textbf {\bibinfo {volume} {106}},\ \bibinfo {pages} {L011701} (\bibinfo {year} {2022})}\BibitemShut {NoStop}%
\bibitem [{\citenamefont {Roy}\ \emph {et~al.}(2023)\citenamefont {Roy}, \citenamefont {Chakrabarti}, \citenamefont {Chavda},\ and\ \citenamefont {Lekala}}]{rhombik_pre}%
  \BibitemOpen
  \bibfield  {author} {\bibinfo {author} {\bibfnamefont {R.}~\bibnamefont {Roy}}, \bibinfo {author} {\bibfnamefont {B.}~\bibnamefont {Chakrabarti}}, \bibinfo {author} {\bibfnamefont {N.~D.}\ \bibnamefont {Chavda}},\ and\ \bibinfo {author} {\bibfnamefont {M.~L.}\ \bibnamefont {Lekala}},\ }\bibfield  {title} {\emph {\bibinfo {title} {Information theoretic measures for interacting bosons in optical lattice}},\ }\href@noop {} {\bibfield  {journal} {\bibinfo  {journal} {Phys. Rev. E}\ }\textbf {\bibinfo {volume} {107}},\ \bibinfo {pages} {024119} (\bibinfo {year} {2023})}\BibitemShut {NoStop}%
\bibitem [{\citenamefont {Baak}\ and\ \citenamefont {Fischer}(2024)}]{fischer_Metrology}%
  \BibitemOpen
  \bibfield  {author} {\bibinfo {author} {\bibfnamefont {J.-G.}\ \bibnamefont {Baak}}\ and\ \bibinfo {author} {\bibfnamefont {U.~R.}\ \bibnamefont {Fischer}},\ }\bibfield  {title} {\emph {\bibinfo {title} {Self-consistent many-body metrology}},\ }\href@noop {} {\bibfield  {journal} {\bibinfo  {journal} {Phys. Rev. Lett.}\ }\textbf {\bibinfo {volume} {132}},\ \bibinfo {pages} {240803} (\bibinfo {year} {2024})}\BibitemShut {NoStop}%
\bibitem [{\citenamefont {Gwak}\ \emph {et~al.}(2021)\citenamefont {Gwak}, \citenamefont {Marchukov},\ and\ \citenamefont {Fischer}}]{GWAK2021168592}%
  \BibitemOpen
  \bibfield  {author} {\bibinfo {author} {\bibfnamefont {Y.}~\bibnamefont {Gwak}}, \bibinfo {author} {\bibfnamefont {O.~V.}\ \bibnamefont {Marchukov}},\ and\ \bibinfo {author} {\bibfnamefont {U.~R.}\ \bibnamefont {Fischer}},\ }\bibfield  {title} {\emph {\bibinfo {title} {Benchmarking the multiconfigurational {Hartree} method by the exact wavefunction of two harmonically trapped bosons with contact interaction}},\ }\href@noop {} {\bibfield  {journal} {\bibinfo  {journal} {Ann. Phys.}\ }\textbf {\bibinfo {volume} {434}},\ \bibinfo {pages} {168592} (\bibinfo {year} {2021})}\BibitemShut {NoStop}%
\bibitem [{\citenamefont {Roy}\ \emph {et~al.}(2019)\citenamefont {Roy}, \citenamefont {L\'ev\^eque}, \citenamefont {Lode}, \citenamefont {Gammal},\ and\ \citenamefont {Chakrabarti}}]{rhombik_quantumreports}%
  \BibitemOpen
  \bibfield  {author} {\bibinfo {author} {\bibfnamefont {R.}~\bibnamefont {Roy}}, \bibinfo {author} {\bibfnamefont {C.}~\bibnamefont {L\'ev\^eque}}, \bibinfo {author} {\bibfnamefont {A.~U.~J.}\ \bibnamefont {Lode}}, \bibinfo {author} {\bibfnamefont {A.}~\bibnamefont {Gammal}},\ and\ \bibinfo {author} {\bibfnamefont {B.}~\bibnamefont {Chakrabarti}},\ }\bibfield  {title} {\emph {\bibinfo {title} {Fidelity and entropy production in quench dynamics of interacting bosons in an optical lattice}},\ }\href@noop {} {\bibfield  {journal} {\bibinfo  {journal} {Quantum Rep.}\ }\textbf {\bibinfo {volume} {1}},\ \bibinfo {pages} {304} (\bibinfo {year} {2019})}\BibitemShut {NoStop}%
\bibitem [{\citenamefont {Molignini}\ \emph {et~al.}(2018)\citenamefont {Molignini}, \citenamefont {Papariello}, \citenamefont {Lode},\ and\ \citenamefont {Chitra}}]{paolo_PhysRevA.98.053620}%
  \BibitemOpen
  \bibfield  {author} {\bibinfo {author} {\bibfnamefont {P.}~\bibnamefont {Molignini}}, \bibinfo {author} {\bibfnamefont {L.}~\bibnamefont {Papariello}}, \bibinfo {author} {\bibfnamefont {A.~U.~J.}\ \bibnamefont {Lode}},\ and\ \bibinfo {author} {\bibfnamefont {R.}~\bibnamefont {Chitra}},\ }\bibfield  {title} {\emph {\bibinfo {title} {Superlattice switching from parametric instabilities in a driven-dissipative {Bose-Einstein} condensate in a cavity}},\ }\href@noop {} {\bibfield  {journal} {\bibinfo  {journal} {Phys. Rev. A}\ }\textbf {\bibinfo {volume} {98}},\ \bibinfo {pages} {053620} (\bibinfo {year} {2018})}\BibitemShut {NoStop}%
\bibitem [{\citenamefont {Lin}\ \emph {et~al.}(2019)\citenamefont {Lin}, \citenamefont {Papariello}, \citenamefont {Molignini}, \citenamefont {Chitra},\ and\ \citenamefont {Lode}}]{paolo_cavity}%
  \BibitemOpen
  \bibfield  {author} {\bibinfo {author} {\bibfnamefont {R.}~\bibnamefont {Lin}}, \bibinfo {author} {\bibfnamefont {L.}~\bibnamefont {Papariello}}, \bibinfo {author} {\bibfnamefont {P.}~\bibnamefont {Molignini}}, \bibinfo {author} {\bibfnamefont {R.}~\bibnamefont {Chitra}},\ and\ \bibinfo {author} {\bibfnamefont {A.~U.~J.}\ \bibnamefont {Lode}},\ }\bibfield  {title} {\emph {\bibinfo {title} {Superfluid--{Mott}-insulator transition of ultracold superradiant bosons in a cavity}},\ }\href@noop {} {\bibfield  {journal} {\bibinfo  {journal} {Phys. Rev. A}\ }\textbf {\bibinfo {volume} {100}},\ \bibinfo {pages} {013611} (\bibinfo {year} {2019})}\BibitemShut {NoStop}%
\bibitem [{\citenamefont {Roy}\ \emph {et~al.}(2024)\citenamefont {Roy}, \citenamefont {Trombettoni},\ and\ \citenamefont {Chakrabarti}}]{rhombik_epjplus}%
  \BibitemOpen
  \bibfield  {author} {\bibinfo {author} {\bibfnamefont {R.}~\bibnamefont {Roy}}, \bibinfo {author} {\bibfnamefont {A.}~\bibnamefont {Trombettoni}},\ and\ \bibinfo {author} {\bibfnamefont {B.}~\bibnamefont {Chakrabarti}},\ }\bibfield  {title} {\emph {\bibinfo {title} {Expansion of strongly interacting dipolar bosons in {1D} optical lattices}},\ }\href@noop {} {\bibfield  {journal} {\bibinfo  {journal} {Eur. Phys. J. Plus}\ }\textbf {\bibinfo {volume} {139}},\ \bibinfo {pages} {831} (\bibinfo {year} {2024})}\BibitemShut {NoStop}%
\bibitem [{\citenamefont {Lin}\ \emph {et~al.}(2021)\citenamefont {Lin}, \citenamefont {Georges}, \citenamefont {Klinder}, \citenamefont {Molignini}, \citenamefont {Büttner}, \citenamefont {Lode}, \citenamefont {Chitra}, \citenamefont {Hemmerich},\ and\ \citenamefont {Keßler}}]{paolo_mott}%
  \BibitemOpen
  \bibfield  {author} {\bibinfo {author} {\bibfnamefont {R.}~\bibnamefont {Lin}}, \bibinfo {author} {\bibfnamefont {C.}~\bibnamefont {Georges}}, \bibinfo {author} {\bibfnamefont {J.}~\bibnamefont {Klinder}}, \bibinfo {author} {\bibfnamefont {P.}~\bibnamefont {Molignini}}, \bibinfo {author} {\bibfnamefont {M.}~\bibnamefont {Büttner}}, \bibinfo {author} {\bibfnamefont {A.~U.~J.}\ \bibnamefont {Lode}}, \bibinfo {author} {\bibfnamefont {R.}~\bibnamefont {Chitra}}, \bibinfo {author} {\bibfnamefont {A.}~\bibnamefont {Hemmerich}},\ and\ \bibinfo {author} {\bibfnamefont {H.}~\bibnamefont {Keßler}},\ }\bibfield  {title} {\emph {\bibinfo {title} {{Mott transition in a cavity-boson system: A quantitative comparison between theory and experiment}}},\ }\href@noop {} {\bibfield  {journal} {\bibinfo  {journal} {SciPost Phys.}\ }\textbf {\bibinfo {volume} {11}},\ \bibinfo {pages} {030} (\bibinfo {year} {2021})}\BibitemShut {NoStop}%
\bibitem [{\citenamefont {Fischer}\ \emph {et~al.}(2015)\citenamefont {Fischer}, \citenamefont {Lode},\ and\ \citenamefont {Chatterjee}}]{fischer_budha}%
  \BibitemOpen
  \bibfield  {author} {\bibinfo {author} {\bibfnamefont {U.~R.}\ \bibnamefont {Fischer}}, \bibinfo {author} {\bibfnamefont {A.~U.~J.}\ \bibnamefont {Lode}},\ and\ \bibinfo {author} {\bibfnamefont {B.}~\bibnamefont {Chatterjee}},\ }\bibfield  {title} {\emph {\bibinfo {title} {Condensate fragmentation as a sensitive measure of the quantum many-body behavior of bosons with long-range interactions}},\ }\href@noop {} {\bibfield  {journal} {\bibinfo  {journal} {Phys. Rev. A}\ }\textbf {\bibinfo {volume} {91}},\ \bibinfo {pages} {063621} (\bibinfo {year} {2015})}\BibitemShut {NoStop}%
\bibitem [{\citenamefont {Penrose}\ and\ \citenamefont {Onsager}(1956)}]{Penrose}%
  \BibitemOpen
  \bibfield  {author} {\bibinfo {author} {\bibfnamefont {O.}~\bibnamefont {Penrose}}\ and\ \bibinfo {author} {\bibfnamefont {L.}~\bibnamefont {Onsager}},\ }\bibfield  {title} {\emph {\bibinfo {title} {{Bose-Einstein} condensation and liquid helium}},\ }\href@noop {} {\bibfield  {journal} {\bibinfo  {journal} {Phys. Rev.}\ }\textbf {\bibinfo {volume} {104}},\ \bibinfo {pages} {576} (\bibinfo {year} {1956})}\BibitemShut {NoStop}%
\bibitem [{\citenamefont {Klaiman}\ and\ \citenamefont {Alon}(2015)}]{variance_ofir2015}%
  \BibitemOpen
  \bibfield  {author} {\bibinfo {author} {\bibfnamefont {S.}~\bibnamefont {Klaiman}}\ and\ \bibinfo {author} {\bibfnamefont {O.~E.}\ \bibnamefont {Alon}},\ }\bibfield  {title} {\emph {\bibinfo {title} {Variance as a sensitive probe of correlations}},\ }\href@noop {} {\bibfield  {journal} {\bibinfo  {journal} {Phys. Rev. A}\ }\textbf {\bibinfo {volume} {91}},\ \bibinfo {pages} {063613} (\bibinfo {year} {2015})}\BibitemShut {NoStop}%
\bibitem [{\citenamefont {Alon}(2019{\natexlab{a}})}]{variance_ofir2019_symmetry}%
  \BibitemOpen
  \bibfield  {author} {\bibinfo {author} {\bibfnamefont {O.~E.}\ \bibnamefont {Alon}},\ }\bibfield  {title} {\emph {\bibinfo {title} {Analysis of a trapped {Bose–Einstein} condensate in terms of position, momentum, and angular-momentum variance}},\ }\href@noop {} {\bibfield  {journal} {\bibinfo  {journal} {Symmetry}\ }\textbf {\bibinfo {volume} {11}} (\bibinfo {year} {2019}{\natexlab{a}})}\BibitemShut {NoStop}%
\bibitem [{\citenamefont {Alon}(2019{\natexlab{b}})}]{variance_ofir2019}%
  \BibitemOpen
  \bibfield  {author} {\bibinfo {author} {\bibfnamefont {O.~E.}\ \bibnamefont {Alon}},\ }\bibfield  {title} {\emph {\bibinfo {title} {Variance of a trapped {Bose-Einstein} condensate}},\ }\href@noop {} {\bibfield  {journal} {\bibinfo  {journal} {J. Phys.: Conf. Ser.}\ }\textbf {\bibinfo {volume} {1206}},\ \bibinfo {pages} {012009} (\bibinfo {year} {2019}{\natexlab{b}})}\BibitemShut {NoStop}%
\bibitem [{\citenamefont {Spekkens}\ and\ \citenamefont {Sipe}(1999)}]{PhysRevA.59.3868}%
  \BibitemOpen
  \bibfield  {author} {\bibinfo {author} {\bibfnamefont {R.~W.}\ \bibnamefont {Spekkens}}\ and\ \bibinfo {author} {\bibfnamefont {J.~E.}\ \bibnamefont {Sipe}},\ }\bibfield  {title} {\emph {\bibinfo {title} {Spatial fragmentation of a {Bose-Einstein} condensate in a double-well potential}},\ }\href@noop {} {\bibfield  {journal} {\bibinfo  {journal} {Phys. Rev. A}\ }\textbf {\bibinfo {volume} {59}},\ \bibinfo {pages} {3868} (\bibinfo {year} {1999})}\BibitemShut {NoStop}%
\bibitem [{\citenamefont {Dutta}\ \emph {et~al.}(2023)\citenamefont {Dutta}, \citenamefont {Lode},\ and\ \citenamefont {Alon}}]{sunayana_scirep}%
  \BibitemOpen
  \bibfield  {author} {\bibinfo {author} {\bibfnamefont {S.}~\bibnamefont {Dutta}}, \bibinfo {author} {\bibfnamefont {A.~U.~J.}\ \bibnamefont {Lode}},\ and\ \bibinfo {author} {\bibfnamefont {O.~E.}\ \bibnamefont {Alon}},\ }\bibfield  {title} {\emph {\bibinfo {title} {Fragmentation and correlations in a rotating {Bose-Einstein} condensate undergoing breakup}},\ }\href@noop {} {\bibfield  {journal} {\bibinfo  {journal} {Sci. Rep.}\ }\textbf {\bibinfo {volume} {13}},\ \bibinfo {pages} {3343} (\bibinfo {year} {2023})}\BibitemShut {NoStop}%
\bibitem [{\citenamefont {Corbo}\ \emph {et~al.}(2017)\citenamefont {Corbo}, \citenamefont {DuBois},\ and\ \citenamefont {Whaley}}]{PhysRevA.96.053627}%
  \BibitemOpen
  \bibfield  {author} {\bibinfo {author} {\bibfnamefont {J.~C.}\ \bibnamefont {Corbo}}, \bibinfo {author} {\bibfnamefont {J.~L.}\ \bibnamefont {DuBois}},\ and\ \bibinfo {author} {\bibfnamefont {K.~B.}\ \bibnamefont {Whaley}},\ }\bibfield  {title} {\emph {\bibinfo {title} {Number-squeezed and fragmented states of strongly interacting bosons in a double well}},\ }\href@noop {} {\bibfield  {journal} {\bibinfo  {journal} {Phys. Rev. A}\ }\textbf {\bibinfo {volume} {96}},\ \bibinfo {pages} {053627} (\bibinfo {year} {2017})}\BibitemShut {NoStop}%
\bibitem [{\citenamefont {Bhowmik}\ and\ \citenamefont {Alon}(2022)}]{anal_2022_dw}%
  \BibitemOpen
  \bibfield  {author} {\bibinfo {author} {\bibfnamefont {A.}~\bibnamefont {Bhowmik}}\ and\ \bibinfo {author} {\bibfnamefont {O.~E.}\ \bibnamefont {Alon}},\ }\bibfield  {title} {\emph {\bibinfo {title} {Longitudinal and transversal resonant tunneling of interacting bosons in a two-dimensional {Josephson} junction}},\ }\href@noop {} {\bibfield  {journal} {\bibinfo  {journal} {Sci. Rep.}\ }\textbf {\bibinfo {volume} {12}},\ \bibinfo {pages} {627} (\bibinfo {year} {2022})}\BibitemShut {NoStop}%
\bibitem [{\citenamefont {Roy}\ \emph {et~al.}(2022)\citenamefont {Roy}, \citenamefont {Chakrabarti},\ and\ \citenamefont {Trombettoni}}]{rhombik_epjd}%
  \BibitemOpen
  \bibfield  {author} {\bibinfo {author} {\bibfnamefont {R.}~\bibnamefont {Roy}}, \bibinfo {author} {\bibfnamefont {B.}~\bibnamefont {Chakrabarti}},\ and\ \bibinfo {author} {\bibfnamefont {A.}~\bibnamefont {Trombettoni}},\ }\bibfield  {title} {\emph {\bibinfo {title} {Quantum dynamics of few dipolar bosons in a double-well potential.}},\ }\href@noop {} {\bibfield  {journal} {\bibinfo  {journal} {Eur. Phys. J. D}\ }\textbf {\bibinfo {volume} {76}},\ \bibinfo {pages} {215303} (\bibinfo {year} {2022})}\BibitemShut {NoStop}%
\bibitem [{\citenamefont {Sakmann}\ \emph {et~al.}(2014)\citenamefont {Sakmann}, \citenamefont {Streltsov}, \citenamefont {Alon},\ and\ \citenamefont {Cederbaum}}]{Universality_of_fragmentation}%
  \BibitemOpen
  \bibfield  {author} {\bibinfo {author} {\bibfnamefont {K.}~\bibnamefont {Sakmann}}, \bibinfo {author} {\bibfnamefont {A.~I.}\ \bibnamefont {Streltsov}}, \bibinfo {author} {\bibfnamefont {O.~E.}\ \bibnamefont {Alon}},\ and\ \bibinfo {author} {\bibfnamefont {L.~S.}\ \bibnamefont {Cederbaum}},\ }\bibfield  {title} {\emph {\bibinfo {title} {Universality of fragmentation in the {Schr\"odinger} dynamics of bosonic {Josephson} junctions}},\ }\href@noop {} {\bibfield  {journal} {\bibinfo  {journal} {Phys. Rev. A}\ }\textbf {\bibinfo {volume} {89}},\ \bibinfo {pages} {023602} (\bibinfo {year} {2014})}\BibitemShut {NoStop}%
\bibitem [{\citenamefont {Z\"ollner}\ \emph {et~al.}(2008)\citenamefont {Z\"ollner}, \citenamefont {Meyer},\ and\ \citenamefont {Schmelcher}}]{PhysRevA.78.013621}%
  \BibitemOpen
  \bibfield  {author} {\bibinfo {author} {\bibfnamefont {S.}~\bibnamefont {Z\"ollner}}, \bibinfo {author} {\bibfnamefont {H.-D.}\ \bibnamefont {Meyer}},\ and\ \bibinfo {author} {\bibfnamefont {P.}~\bibnamefont {Schmelcher}},\ }\bibfield  {title} {\emph {\bibinfo {title} {Tunneling dynamics of a few bosons in a double well}},\ }\href@noop {} {\bibfield  {journal} {\bibinfo  {journal} {Phys. Rev. A}\ }\textbf {\bibinfo {volume} {78}},\ \bibinfo {pages} {013621} (\bibinfo {year} {2008})}\BibitemShut {NoStop}%
\bibitem [{\citenamefont {Hunn}\ \emph {et~al.}(2013)\citenamefont {Hunn}, \citenamefont {Zimmermann}, \citenamefont {Hiller},\ and\ \citenamefont {Buchleitner}}]{PhysRevA.87.043626}%
  \BibitemOpen
  \bibfield  {author} {\bibinfo {author} {\bibfnamefont {S.}~\bibnamefont {Hunn}}, \bibinfo {author} {\bibfnamefont {K.}~\bibnamefont {Zimmermann}}, \bibinfo {author} {\bibfnamefont {M.}~\bibnamefont {Hiller}},\ and\ \bibinfo {author} {\bibfnamefont {A.}~\bibnamefont {Buchleitner}},\ }\bibfield  {title} {\emph {\bibinfo {title} {Tunneling decay of two interacting bosons in an asymmetric double-well potential: A spectral approach}},\ }\href@noop {} {\bibfield  {journal} {\bibinfo  {journal} {Phys. Rev. A}\ }\textbf {\bibinfo {volume} {87}},\ \bibinfo {pages} {043626} (\bibinfo {year} {2013})}\BibitemShut {NoStop}%
\bibitem [{\citenamefont {Juli\'a-D\'{\i}az}\ \emph {et~al.}(2010)\citenamefont {Juli\'a-D\'{\i}az}, \citenamefont {Martorell}, \citenamefont {Mel\'e-Messeguer},\ and\ \citenamefont {Polls}}]{PhysRevA.82.063626}%
  \BibitemOpen
  \bibfield  {author} {\bibinfo {author} {\bibfnamefont {B.}~\bibnamefont {Juli\'a-D\'{\i}az}}, \bibinfo {author} {\bibfnamefont {J.}~\bibnamefont {Martorell}}, \bibinfo {author} {\bibfnamefont {M.}~\bibnamefont {Mel\'e-Messeguer}},\ and\ \bibinfo {author} {\bibfnamefont {A.}~\bibnamefont {Polls}},\ }\bibfield  {title} {\emph {\bibinfo {title} {Beyond standard two-mode dynamics in bosonic {Josephson} junctions}},\ }\href@noop {} {\bibfield  {journal} {\bibinfo  {journal} {Phys. Rev. A}\ }\textbf {\bibinfo {volume} {82}},\ \bibinfo {pages} {063626} (\bibinfo {year} {2010})}\BibitemShut {NoStop}%
\end{thebibliography}%
\end{document}